\documentclass[preprint]{aastex631}

\newcommand{\red}{\color{black}}
\newcommand{\black}{\color{black}}
\newcommand{\de}{$^{\circ}$}
\newcommand{\Rs}{ R$_{\odot}$}
\newcommand{\pBk}{\textit{pB}$_k$}
\newcommand{\Bs}{$B_{\odot}$}

\shortauthors{Edwards et al.}

\graphicspath{{./}{figures/}}

\begin{document}

\title{Derived electron densities from linear polarization observations of the visible-light corona during the 14 December 2020 total solar eclipse}


\correspondingauthor{Liam T. Edwards}
\email{lie6@aber.ac.uk}

\author[0000-0002-9222-8648]{Liam T. Edwards}
\affiliation{Department of Physics \\
Aberystwyth University \\
Ceredigion, Cymru, SY23 3BZ}

\author{Kaine A. Bunting}
\affiliation{Department of Physics \\
Aberystwyth University \\
Ceredigion, Cymru, SY23 3BZ}

\author[0000-0002-6845-1698]{Brad Ramsey}
\affiliation{Department of Physics \\
Aberystwyth University \\
Ceredigion, Cymru, SY23 3BZ}

\author[0000-0002-1366-678X]{Matthew Gunn}
\affiliation{Department of Physics \\
Aberystwyth University \\
Ceredigion, Cymru, SY23 3BZ}

\author[0000-0002-0500-5789]{Tomos Fearn}
\affiliation{Department of Computer Science \\
Aberystwyth University \\
Ceredigion, Cymru, SY23 3DB}

\author{Thomas Knight}
\affiliation{Department of Physics \\
Aberystwyth University \\
Ceredigion, Cymru, SY23 3BZ}

\author[0000-0003-0581-1278]{Gabriel Domingo Muro}
\affiliation{Department of Physics \\
Aberystwyth University \\
Ceredigion, Cymru, SY23 3BZ}
\affiliation{Space Radiation Lab, \\
California Institute of Technology, \\
Pasadena, CA, USA 91125}

\author[0000-0002-6547-5838]{Huw Morgan}
\affiliation{Department of Physics \\
Aberystwyth University \\
Ceredigion, Cymru, SY23 3BZ}

\begin{abstract}
A \red new \black instrument was designed to take visible-light (VL) polarized brightness ($pB$) observations of the solar corona during the 14 December 2020 total solar eclipse. The instrument, called the Coronal Imaging Polarizer (CIP), consisted of a 16 MP CMOS detector, a linear polarizer housed within a piezoelectric rotation mount, and an f-5.6, 200 mm DSLR lens. Observations were successfully obtained, despite poor weather conditions, for five different exposure times (0.001 s, 0.01 s, 0.1 s, 1 s, and 3 s) at six different orientation angles of the linear polarizer (0\de, 30\de, 60\de, 90\de, 120\de, and 150\de). The images were manually aligned using the drift of background stars in the sky and images of different exposure times were combined using a simple \red signal-to-noise ratio cut. The polarization and brightness of the local sky is also estimated and the observations were subsequently corrected. \black The $pB$ of the K-corona \red was determined \black using least squares fitting and radiometric calibration was done relative to the Mauna Loa Solar Observatory (MLSO) K-Cor $pB$ observations from the day of the eclipse. The $pB$ data was then inverted to acquire the coronal electron density, $n_e$, \red for an equatorial streamer and a polar coronal hole\black, which agreed very well with previous studies. \red The effect of changing the number of polarizer angles used to compute the $pB$ is also discussed and it is found that the results vary by up to $\sim$ 13\% when using all six polarizer angles versus only a select three angles.\black
\end{abstract}


\section{Introduction} \label{sec:introduction}
\red A total solar eclipses (TSE) provides a unique opportunity to observe the visible-light (VL) corona down to the solar limb for a few minutes during totality, and allow continuous observations from the limb out to several solar radii. \black Not only does the Moon block out the solar disk, but it also lowers the local sky brightness along the eclipse path which makes it perfect for observing the significantly fainter corona \citep{lang2010}. This lower coronal region ($<$ 2 \Rs) is the origin of the solar wind and is therefore crucial to observe in order to better understand how it is formed and accelerates to supersonic speeds \citep{2020JPhCS1620a2006H, 2017BAMS...98.2593S, 2007RvGeo..45.1004M}. TSEs have been used to obtain valuable information about a variety of solar phenomena, including coronal streamers \citep{2022SoPh..297...28P}, coronal mass ejections (CMEs) (\cite{2021ApJ...914L..39B, 2020SoPh..295...24F, 2004A&A...420..709K}), coronal holes \citep{2022SoPh..297...28P}, plasma flows \citep{2014ApJ...797...10S, 2009ApJ...701L...1D}, prominences \citep{2014SoPh..289.2487J}, and coronal jets \citep{2018ApJ...860..142H}. Physical properties of the coronal plasma have also been studied extensively during TSEs, typically, temperature, density, and velocity \citep{2023ApJS..265...11D, 2023gabe, 2020ApJ...904..178B, 2014SoPh..289.2021R, 2011ApJ...734..120H, 2010ApJ...708.1650H, 2009SPD....40.1403R, 2007ApJ...663..598H}.

A coronagraph \red is required in order to observe the VL corona outside of a total solar eclipse\black. First designed by Bernard Lyot in the 1930's \citep{1933JRASC..27..225L}, it consists of a telescope with an opaque disk that is positioned to block out the bright disk of the Sun. This is crucial because the photosphere is $\sim$ 10$^6$ times brighter than the corona itself. In fact, the Earth's sky is also much brighter than the corona - of the order $\sim$ 10$^5$ times brighter at 20 \Rs \, - therefore, any ground-based coronagraph, such as \red the \black COronal Solar Magnetism Observatory's (COSMO) K-Coronagraph (K-Cor; \citet{2013ApJ...774...85H}), \red is affected by the Earth's own atmosphere\black. To overcome this issue, several space-based coronagraphs have been launched in recent decades, for example, the Large Angle Spectrometric Coronagraph (LASCO; \citet{1995SoPh..162..357B}) onboard the Solar and Heliospheric Observatory (SOHO; \citet{1995SoPh..162....1D}) and the Sun-Earth Connection Coronal and Heliospheric Investigation's (SECCHI; \citet{2008SSRv..136...67H}) COR1/2 instruments onboard the Solar Terrestrial Relations Observatory (STEREO; \citet{2008SSRv..136....5K}). However, despite the unprecedented access to the corona they have given the field of solar physics, there is still a fundamental issue for any type of coronagraph to overcome which is the stray light resulting from the diffraction of incoming light at the occulter edge. In order to mitigate this, most occulters will block not only the solar disk but also a portion of the very lower solar corona - out to around 1.5\Rs \, for internally occulted coronagraphs \citep{2008JOSAA..25..182V}. One way to circumvent this issue is to increase the distance between the detector and the occulter, which is possible to achieve in space by the use of formation flying cube satellites (e.g., ASPIICS; \citet{2010SPIE.7731E..18L})\red, however, these type of instruments are still in their infancy. As a result, observations of the inner corona during a TSE are unmatched in the VL regime. \black

The VL corona is composed of light from several different sources, primarily from the scattering of photospheric light by free electrons in the corona and dust in the interplanetary plane, termed the K- and F-corona, respectively. In the case of the K-corona, the scattering process - known as Thomson scattering - produces a strongly tangentially polarized component to the total VL brightness (for an in-depth overview of this mechanism see \citet{2015arXiv151200651I}), whereas the F-corona is considered to be unpolarized below $\sim$ 3 \Rs \, \citep{2007A&A...471L..47M}. As a result, observing the VL corona below this height with a linear polarizer during a TSE can be considered to be a measurement of the K-coronal brightness only. There are also other types of brightness contributions to the total coronal brightness, namely the E-corona (emission) which consists of spectral line emission from highly ionized atoms, but these are considered to be negligible for the purposes of this work. The K-coronal brightness component, $B_K$, represents the structure and amount of coronal plasma irrespective of its temperature, unlike EUV or x-ray observations. The K-coronal component of the polarized brightness (\pBk) gives the electron density ($n_e$) which is calculated using an inversion method first developed by \citet{1950BAN....11..135V} and later improved upon by \citet{2001ApJ...548.1081H} and \citet{2002A&A...393..295Q}. Equation \ref{eq:density_inversion} describes the relationship between the polarized brightness of the K-corona and the electron density which is used to acquire the electron density from the TSE images:

\begin{equation}
    \label{eq:density_inversion}
    pB_k \propto \int_{LOS} n_e(r) \cdot G(s, \rho) \, ds
\end{equation}

where $G$ is a geometrical weighting function, $\rho$ is the distance between the Sun and the intercept between the line-of-sight (LOS) and the plane-of-sky (POS), and $s$ is the distance between the intercept point on the POS and an arbitrary point in the corona along the LOS (see Figure 1 in \citet{2002A&A...393..295Q} for more detail). This is a well-established technique and several studies have used this inversion of $pB$ to obtain coronal electron densities (e.g. \cite{2022MNRAS.tmp.2971L, 2020ApJ...904..178B, 2012SoPh..277..267S, 2001ApJ...548.1081H, 1986BASI...14..217R, 1977SoPh...55..121S}). Several studies have also been conducted attempting to separate the K- and F-components of the solar corona (e.g. \cite{2021AGUFMSH15D2057B, 2009Ge&Ae..49..830F, 2007A&A...471L..47M, 1982A&A...112..241D, 1972ApJ...176..497C}). Most of the previous studies involving TSE observations require some level of image processing as a result of several factors - primarily the sharp decrease in brightness with radial height from the Sun. For example, since the Moon and the Sun move relative to each other with respect to the background stars, all TSE images need to be coaligned and there are a number of different methods which can be used to do this. One of the most common of these methods over the past decade or so has been to use a modified phase correlation technique by \cite{2009ApJ...706.1605D}, whereas others have used more manual methods such as using the drift of background stars as they move relative to the TSE \citep{2020ApJ...904..178B}. There are also many different image processing techniques that have been developed to better reveal various structures and phenomena (\cite{2022SoPh..297...27P, 2020NewA...7901383Q, 2014SoPh..289.2945M, 2013ApJS..207...25D, 2012ApJ...752..145B, 2011ApJ...737...88D}). The outline of this paper is as follows: section \ref{subsec:eclipse} describes the 2020 TSE in more detail, section \ref{sec:method} summarises the design of the instrument (\ref{subsec:CIP}), and the calibrating, processing, coaligning, and inversion to derive the coronal electron densities (\ref{subsec:alignment} - \ref{subsec:kcor}). Section \ref{sec:results} presents the results of the study, with sections \ref{sec:discussion} and \ref{sec:conclusion} disseminating and concluding the results of this work, respectively.

\subsection{14 December 2020 Total Solar Eclipse}\label{subsec:eclipse}
\vspace{2mm}

\begin{figure}[h!]
    \centering
    \includegraphics[width=0.75\textwidth]{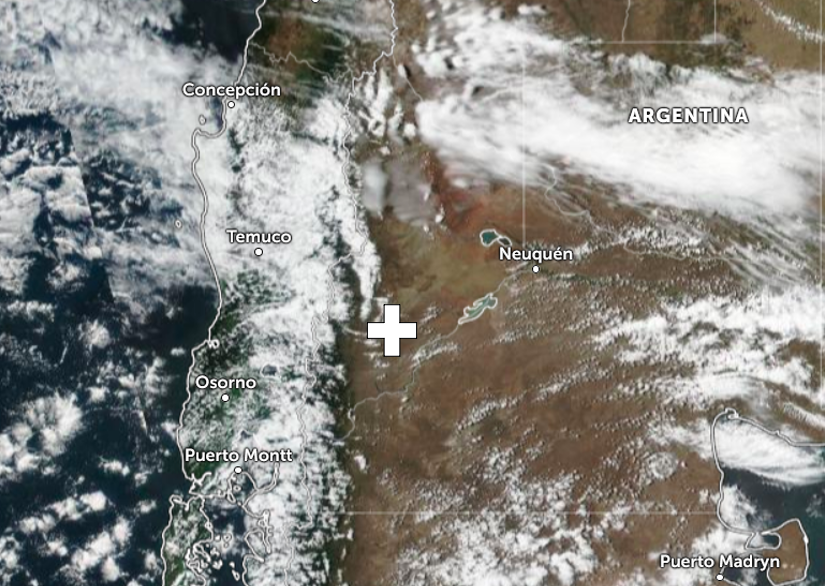}
    \caption{\red Satellite image of cloud cover above the observation site in Neuqu\'{e}n province, denoted by the white marker, taken at approximately 13:30 local time (Credit: Zoom Earth)}
    \label{fig:eclipse map}
\end{figure}

The TSE on 14 December 2020 was observed by a team from Aberystwyth University at a site in Neuqu\'{e}n province, Argentina. The observation site was located at 39\de 42' 40.4" S, 70\de 23' 57.6" W, and an altitude of $\approx$ 1,082 m (3,550 ft). Totality lasted $\approx$ 129 seconds with the time of maximum eclipse at 13:07:58 local time (16:07:58 UTC) and the apparent altitude of the Sun above the horizon was $\approx$ 75\de. The weather conditions at the time of observation were not optimal with intermittent cloud cover and very strong gusts of up to 70 km/h which resulted in a lot of airborne dust. Furthermore, a small, wispy cloud passed across the Sun's disk for $\approx$ 12 seconds. This had an adverse effect on the quality of the data but, despite the conditions, the data captured during totality was still usable. \red Figure \ref{fig:eclipse map} shows the location of the observation site (white marker) along with the cloud cover at approximately the time of the eclipse\black. During this eclipse, a CME had erupted from the eastern limb of the Sun $\approx$ 110 minutes before totality, providing a truly unique opportunity to study CME dynamics right down to the solar limb. The LASCO CME catalog \citep{2009EM&P..104..295G} states that the CME first appeared in the C2 field-of-view (FOV) at 15:12:10 UT with an estimated linear speed of 437 km/s, mass of 3 $\times$ 10$^{12}$ kg, and a central position angle of 121\de. The CME is discussed in detail by \cite{2021ApJ...914L..39B}.

\black
\newpage

\section{Instrument Design and Data Processing}
\label{sec:method}

\vspace{2mm}
\subsection{The Coronal Imaging Polarizer (CIP)}\label{subsec:CIP}
\vspace{2mm}

\begin{figure}[h!]
    \centering
    \includegraphics[width=0.99\textwidth]{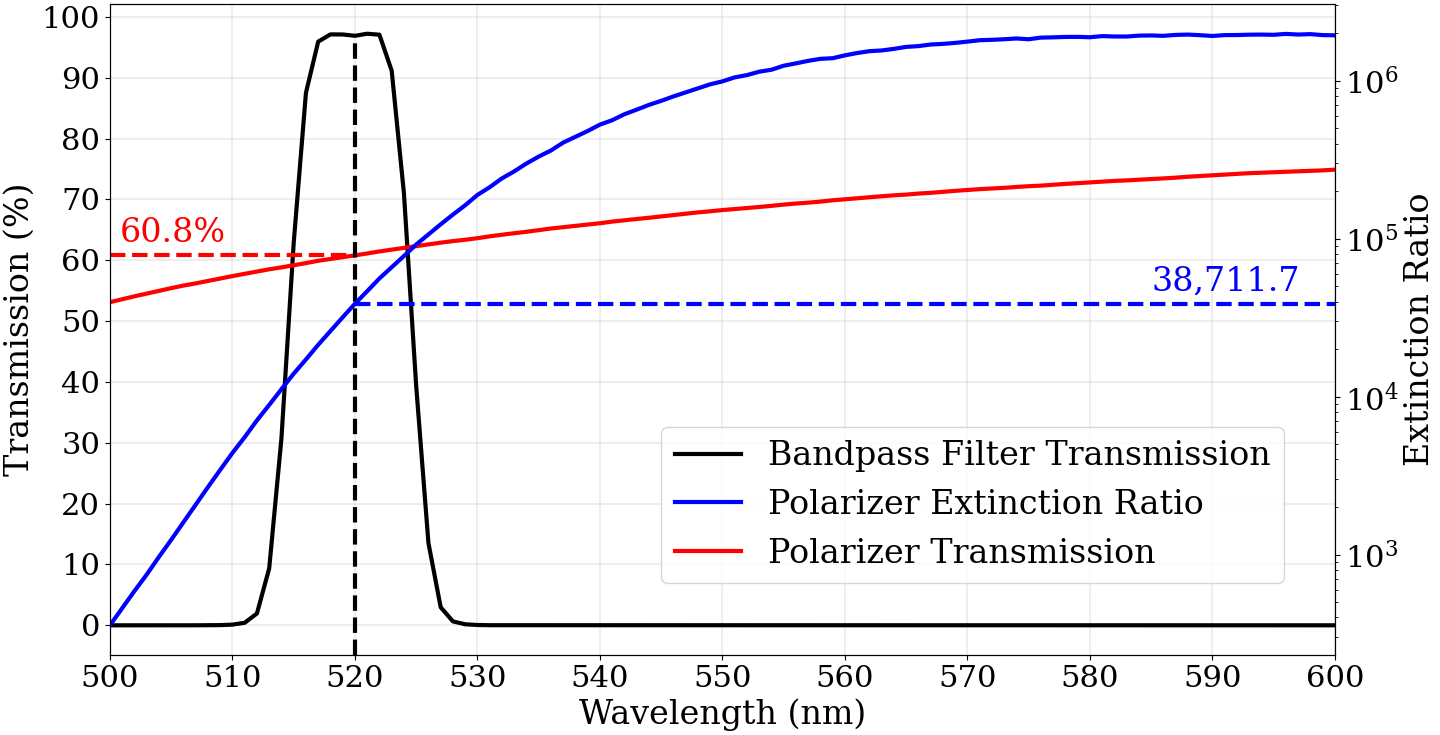}
    \caption{\black Extinction ratio and transmissions for both the band-pass filter and linear polarizer \red measured by their respective manufacturers. \black The dotted horizontal lines correspond to the polarizer's extinction ratio \red (blue) \black and transmission \red (red) \black at the band-pass filter's center wavelength (Data from Thorlabs)}
    \label{fig:transmissions graph}
\end{figure}

The instrument used to observe the corona during the total solar eclipse \red was \black a VL \red linear \black polarization imager designed and built at Aberystwyth University. The Coronal Imaging Polarizer (CIP) \red was \black designed to be relatively cheap and simple to build, easy to assemble, and lightweight in order to be able to relocate quickly on the day of an eclipse if necessary. It \red consisted \black of an objective lens, a VL band-pass filter\footnote{\url{https://www.thorlabs.com/thorproduct.cfm?partnumber=FBH520-10}}, a \red linear \black polarizer\footnote{\url{https://www.thorlabs.com/thorproduct.cfm?partnumber=LPVISC100}} housed in a rotating mount\footnote{\url{https://www.thorlabs.com/newgrouppage9.cfm?objectgroup\_id=12829}}, and a CMOS sensor\footnote{\url{https://www.atik-cameras.com/product/atik-horizon-ii/}}. The objective lens, an f-5.6, 200 mm focal length DSLR lens, can be adjusted to give different fields of view if required. The VL band-pass filter has a center wavelength of 520 nm (denoted by the vertical dashed black line in Figure \ref{fig:transmissions graph}) where transmission is $>$ 90\%, and a bandwidth (FWHM) of 10 nm. Since the peak of the Sun's emission is around 500 nm, the filter is well-placed to collect the maximum amount of light possible. 

\black 

Traditionally, when taking VL polarization observations of the corona, a polarizer is either manually rotated through a set number of polarization angles (typically 3 - 5) or several different instruments are set up, each designed to capture the light of a single \red polarizer orientation angle. \black In contrast, CIP used a piezoelectrically-driven rotation mount from Thorlabs to automatically rotate the polarizer through six different polarization angles (0\de, 30\de, 60\de, 90\de, 120\de, and 150\de). The motorized rotation mount allows for 360\de \, rotation with a maximum rotation speed of 430\de/second and has an accuracy of $\pm$ 0.4\de, resulting in high precision and fast rotation. The energy requirement of the motor is very low, needing only a maximum of 5.5 V DC input with a typical current consumption of 800 and 50 mA during movement and standby, respectively.

The camera used for CIP was the Horizon II - a 16 mega-pixel CMOS camera developed by Atik. It uses the Panasonic MN34230 4/3" CMOS sensor with a 4644 x 3506 resolution and has a very low readout noise ($\sim$ 1 e$^-$). It is powered by a 12 V 2 A DC input and has a minimum exposure time of 18 $\mu$s and unlimited maximum exposure. It is also cooled via an internal fan and can maintain a $\Delta$T of -40\de C, meaning the camera can still maintain low thermal noise even at high ambient background temperatures. It is a black-and-white camera since using an RGB camera requires extra steps when processing the data (e.g. demosaicking), and the correction and alignment must be done individually for each RGB channel - see \citet{2020ApJ...904..178B}, for example. In order for the instrument to run in the most efficient way possible, the data collection process was fully automated. The rotation mount rotates the polarizer to a specific angle, then the camera collects a sequence of images of varying exposure times from 0.001 - 3 seconds, then the polarizer is rotated to the next angle, and the camera would run through the same sequence as before, and so on for all six polarization angles. The time taken for one complete cycle of data collection is around 30 seconds which resulted in two full data sets and a third partial data set.

\begin{figure}[h!]
    \centering
    \includegraphics[width=0.99\textwidth]{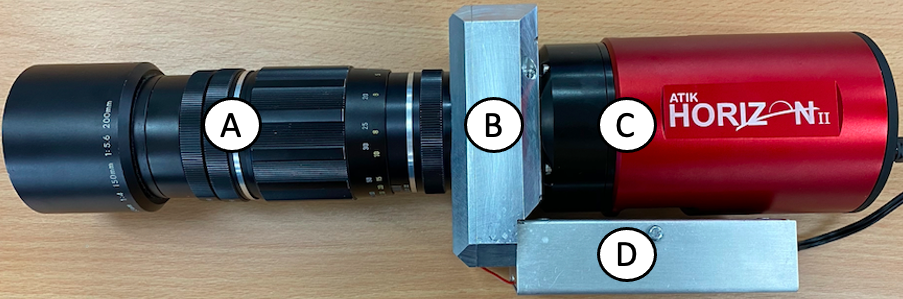}
    \caption{The Coronal Imaging Polarizer (CIP) - \textit{\textbf{A:}} f-5.6, 200 mm objective lens. \textit{\textbf{B:}} housing containing the rotation mount, polarizer, and VL filter. \textit{\textbf{C:}} Atik Horizon II 16 MP CMOS camera \textit{\textbf{D:}} housing for the interface board of the rotation mount}
    \label{fig:CIP}
\end{figure}

\newpage

\subsection{Data Processing}

\subsubsection{Linearity}\label{subsec:linearity}
\vspace{2mm}

\begin{figure}[h!]
    \centering
    \includegraphics[width=0.99\textwidth]{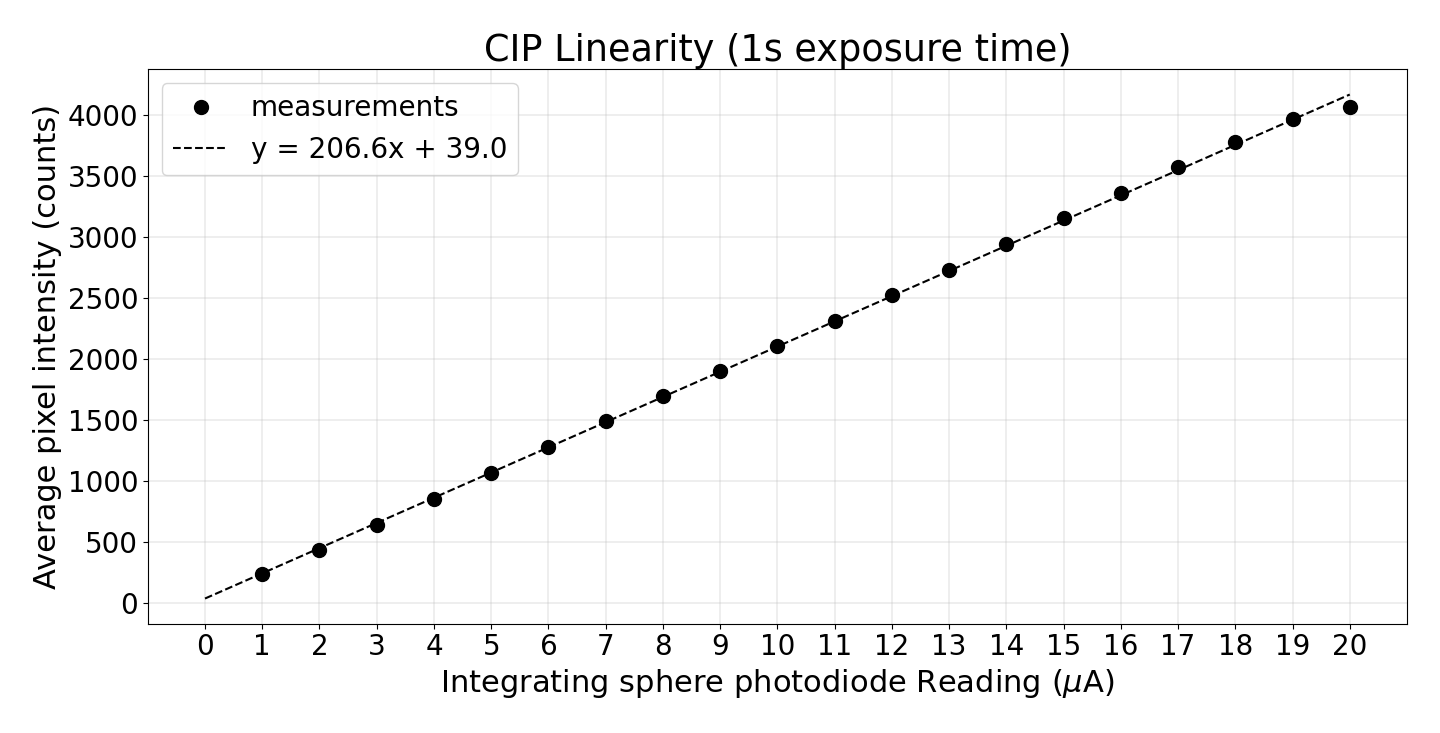}
    \caption{Average pixel intensity as a function of the integrating sphere's photodiode reading (black dots) and the calculated line of best fit (dashed line) giving an R$^2$ value of 0.99956}
    \label{fig:CIP linearity}
\end{figure}

The response of the imaging sensor to incoming light intensity is measured by using an integrating sphere, where the \red level of illumination in the sphere is measured by a photodiode (in Amps). The integrating sphere is not calibrated to give the light levels in SI units but it is linear so if the photodiode current doubles the light level in the sphere is doubled. The illumination of the sphere was gradually increased by 1 $\mu$A until it reached the saturation point. \black Five measurements were taken and an average intensity was calculated for each photodiode reading. Figure \ref{fig:CIP linearity} presents the results of this linearity test and it clearly shows that the sensor used in CIP has a linear response. The linearity begins to break down close to the saturation limit ($\approx$ 4096 counts) which is to be expected and is accounted for when the images taken at different exposure times are combined for each polarization angle (Section \ref{subsec:combination}). The linearity of the sensor was quantified using equation (\ref{eq:linearity}):

\begin{equation}\label{eq:linearity}
    Linearity \, (\%) = \frac{(MPD + MND)}{MI} \times 100
\end{equation}

where MPD, MND, and MI are the maximum positive deviation, maximum negative deviation, and maximum intensity, respectively. The maximum positive and negative deviations are found from the line of best fit. This calculation was performed twice - once including the last data point where the linearity begins to break down, and a second time without including the aforementioned data point. These calculations result in linearities of 2.99\% and 0.96\%, respectively. Both of these values are excellent linearities since most CMOS sensors have linearities on the order of several percent \citep{linearityReferencePaper}.

\newpage

\subsubsection{Flat field and dark frame correction}\label{subsec:flat/dark correction}
\vspace{2mm}

\begin{figure}[h!]
    \centering
    \includegraphics[width=0.74\textwidth]{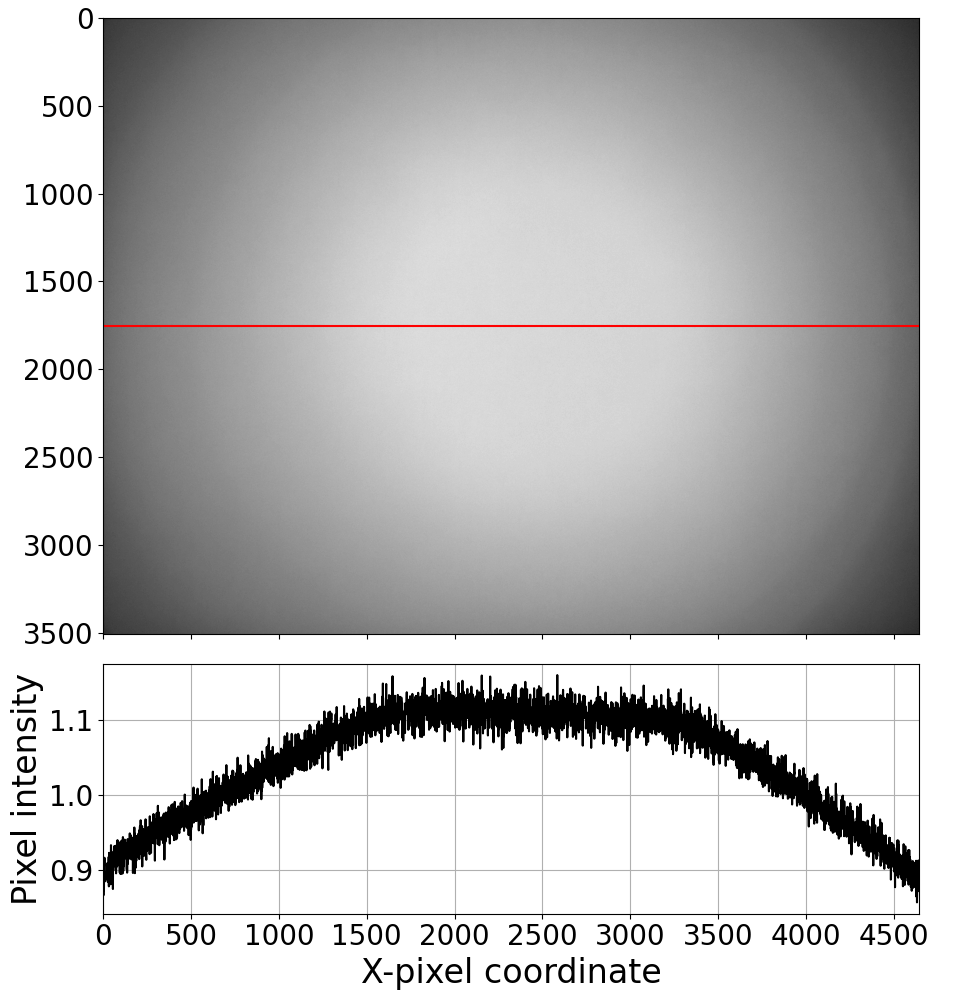}
    \caption{\black Master flat field image for a polarization angle of 0\de \, (\textit{top}) and a \red normalized \black intensity profile taken at the midpoint of the image shown by the red horizontal line (\textit{bottom})}
    \label{fig:flat field line profile}
\end{figure}

\red Five flat field images were taken at an exposure time of 0.01s for each polarization angle using an integrating sphere. \black A master flat field image was then produced for each polarization angle using the \red open-source plugin AstroImageJ \black - an example of one of these is seen in Figure \ref{fig:flat field line profile} along with an intensity profile taken at the midpoint of the image.

\black

\begin{figure}[h!]
    \centering
    \includegraphics[width=0.99\textwidth]{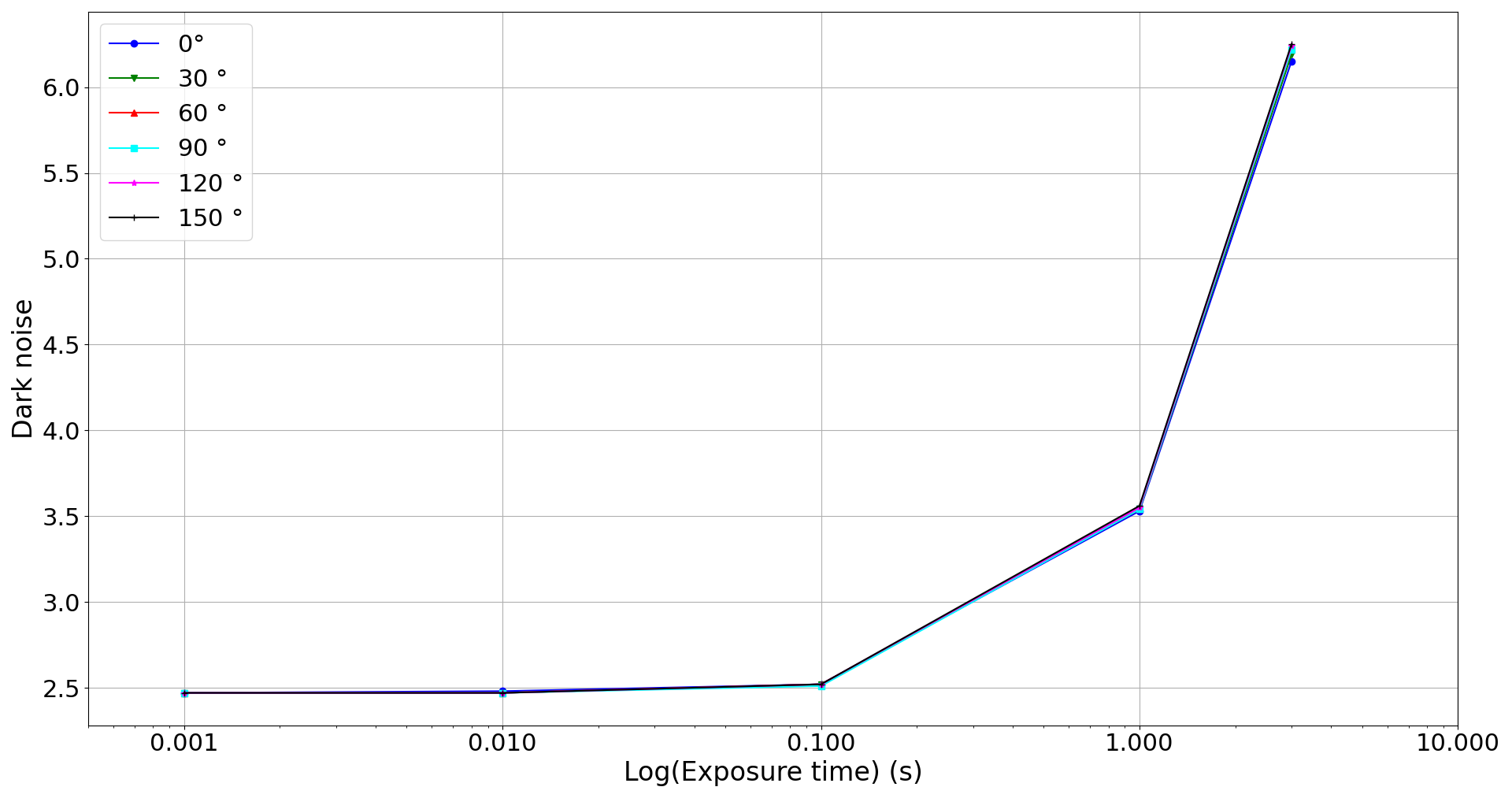}
    \caption{\red Average dark noise for all polarization angles as a function of the exposure time \black}
    \label{fig:dark noise}
\end{figure}

Two different types of dark field images were attempted: firstly, setting the exposure time of the camera to zero and taking several pictures to get an average; and secondly, blocking the front of the instrument with the lens cap and running the full eclipse sequence \red five \black times. Unfortunately, the Atik Horizon II has a minimum exposure time of 18 $\mu$s so the first method was not able to be done. For the second method, the instrument was taken to a dark room with the lens cap taped over the lens. Several dark frames were taken in this way for each polarization angle and exposure time used in the eclipse sequence. It is expected that the mean values of the intensity (along with their standard deviations) should be similar for each polarization angle and exposure time and this is indeed what was seen. Figure \ref{fig:dark noise} shows the dark noise, $\sigma_{dark}$, calculated by taking the average standard deviation of all dark frames taken at each exposure time and polarization angle. It is clear that pixels in the detector follow the same pattern and behave uniformly across all polarization angles.

Once the master flat field and dark frame images were produced, the raw eclipse images were calibrated using equation (\ref{eq:flat_dark_calibration_equation}):

\begin{equation}
    \label{eq:flat_dark_calibration_equation}
    C_{\theta,t} = \frac{(R_{\theta,t} - D_t)*m}{(F_{\theta} - D_t)}
\end{equation}

where $C_{\theta,t}$ is the \red reduced \black image, $R_{\theta,t}$ is the raw image taken at a specific polarization angle ($\theta$) and exposure time ($t$), $D_t$ is the dark frame for that particular exposure, $F_{\theta}$ is the flat-field for that particular polarization angle, and $m$ is the image-averaged value of $(F_{\theta} - D_t)$. At this stage, the scale of the image was calculated. The radius of the Moon was found to be 255.5 $\pm$ 0.5 pixels by fitting a circle to points plotted along the lunar limb using ImageJ. One of the lowest exposure images was used to obtain this radius in order to reduce the brightness from the lower corona to obtain the most accurate value possible. The Moon's apparent radius at the time of the eclipse was found to be \red 1,000.145" using the Stellarium software \citep{Zotti_Hoffmann_Wolf_Chéreau_Chéreau_2021}\black, therefore, the full-resolution images provided a spatial resolution of \red 3.92 \black $\pm$ 0.01 arcsec/pixel. 

\newpage
\subsubsection{Image coalignment}\label{subsec:alignment}
\vspace{2mm}

\begin{figure}[h!]
    \centering
    \includegraphics[width=0.49\textwidth]{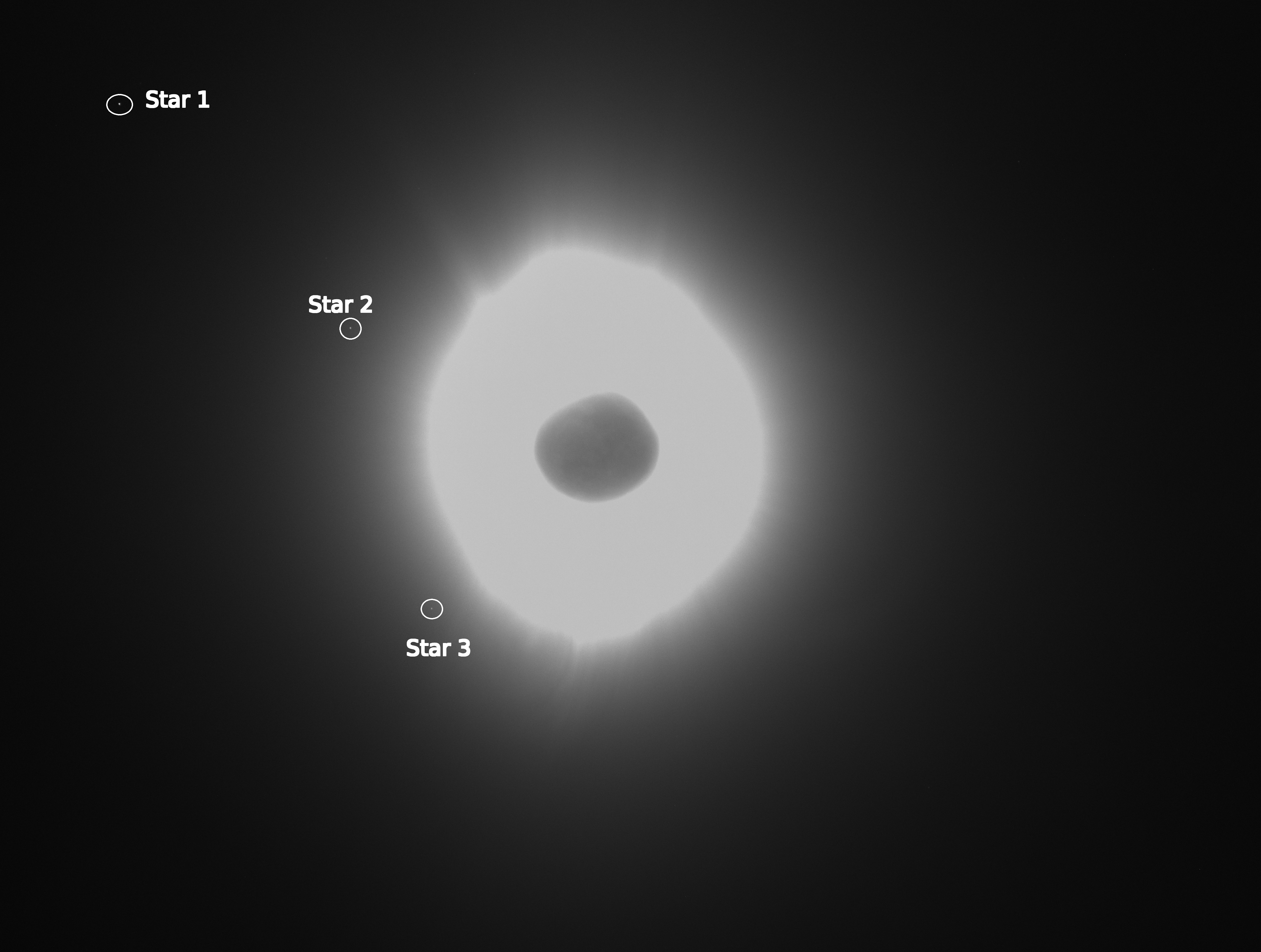}
    \hfill
    \includegraphics[width=0.49\textwidth]{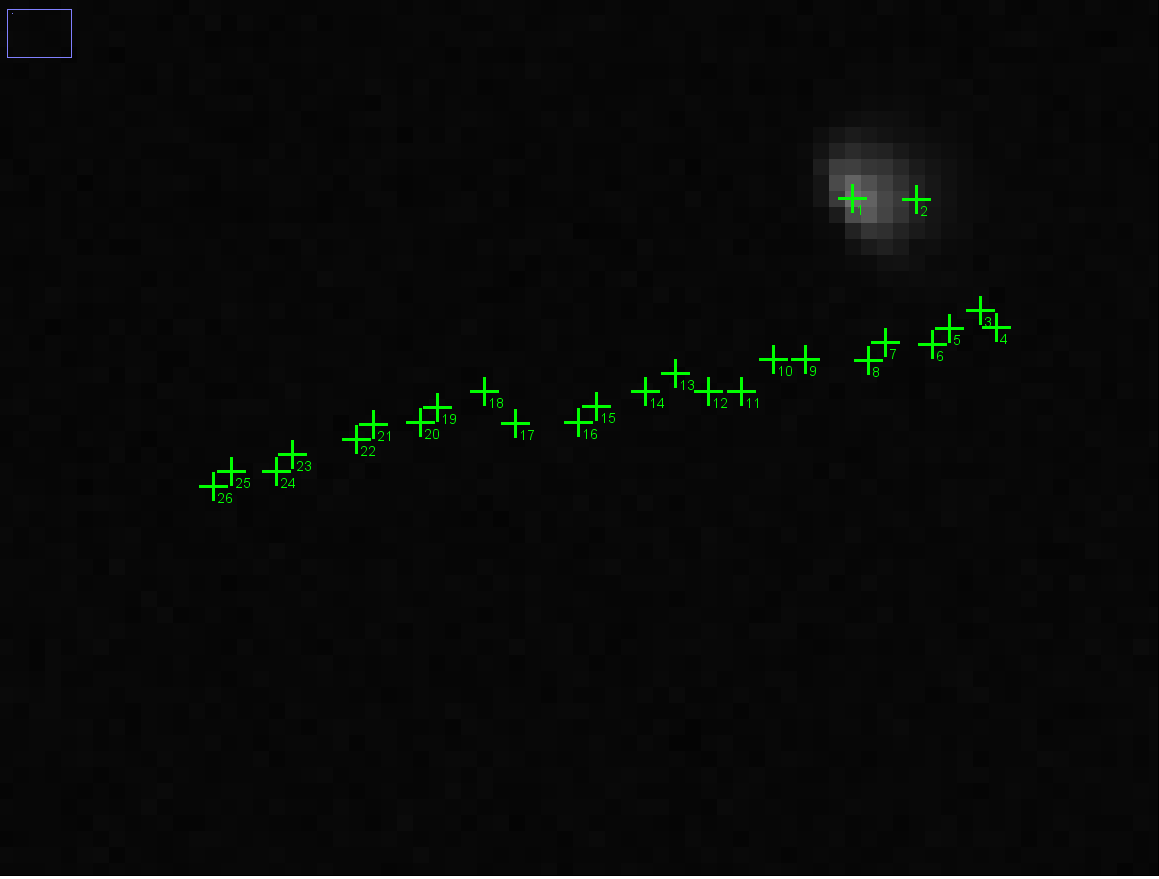}
    \caption{\textit{Left:} Locations of the three visible stars in a 3 s exposure time image. \textit{Right:} Locations of the brightest pixel in star 1 for all 1 and 3 s exposures}
    \label{fig:star locations and drift}
\end{figure}

\begin{figure}[h!]
    \centering
    \includegraphics[width=0.99\textwidth]{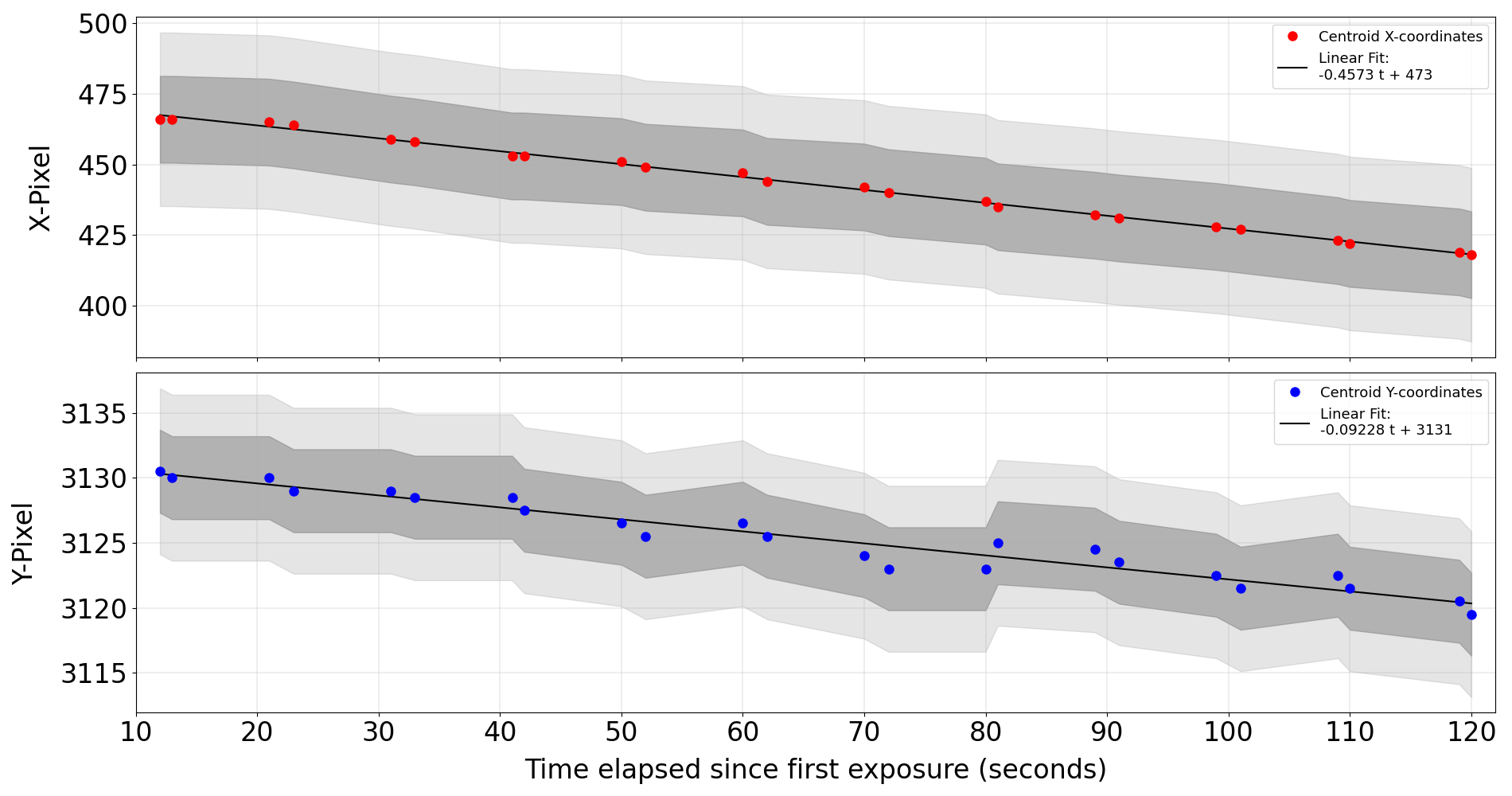}
    \caption{Pixel drifts in both the x- and y-coordinates for star 1. The data points represent the pixel coordinate of the brightest pixel during the 1 and 3 s exposures. The darker and lighter shaded regions show one and two standard deviations, respectively}
    \label{fig:pixel drifts}
\end{figure}

After the initial calibration of the raw images with the dark and flat frames, they \red were then \black coaligned before the images with different exposure times \red could \black be combined. This step is crucial because, although a motorized tracking mount was used to track the solar center, there are still some factors that need to be accounted for and corrected. The two main factors are the high winds at ground level throughout the eclipse and the fact that the Moon is moving with respect to the Sun. The effect of the wind on the tracking mount can be seen in Figures \ref{fig:star locations and drift} and \ref{fig:pixel drifts}. As discussed in section \ref{sec:introduction}, there are several different methods of coaligning eclipse images such as the use of phase correlation (e.g. \citet{2009ApJ...706.1605D}) or by using the positions of stars visible in the exposures (e.g. \citet{2020ApJ...904..178B}). In this study, three stars were visible in the images taken at longer exposure times (see Figure \ref{fig:star locations and drift}) and were identified using Stellarium to be HD 157056 (star 1), HD 157792 (star 2), and HD 158643 (star 3), respectively. The images were initially coaligned using the brightest of these stars (HD 157056), which is star 1 in Figure \ref{fig:star locations and drift}, since its drift in both the x- and y-direction was fairly consistent (Figures \ref{fig:star locations and drift} and \ref{fig:pixel drifts}). \red For the shorter exposure images, where the stars were not visible, it was assumed that the star did not move much relative to its position in the 1 and 3 s exposures. \black Shifts in both the x- and y-pixels were calculated for each \red image, based on their respective exposure times, \black using the equations found in Figure \ref{fig:pixel drifts} and all \red of the images were \black co-aligned accordingly. The images are then re-binned through 8 $\times$ 8 pixel averaging to improve the signal quality\red, which reduced the resolution of the images to 31.25 $\pm$ 0.01 arcsec/pixel. \black 

\subsubsection{Image combination}\label{subsec:combination}

The next step was to combine all the \red images of different exposure times \black for each individual polarization angle. Firstly, all negative pixel values were set to zero and the images were normalized by their respective exposure times (DN/s) using equation \ref{eq:count rate}:

\begin{equation} \label{eq:count rate}
    I_{p}' = \frac{I_p}{\Delta t_e}
\end{equation}

where $I_{p}'$ is the pixel intensity ($I_p$) for each polarization angle $p$, normalised by exposure time ($\Delta t_e$). Since the eclipse images have a very high dynamic range between the brightest and darkest pixels, they need to be \red combined together in a way that takes the pixel value itself into account. In the shorter exposure images, the inner coronal signal is strong but the outer coronal signal is too weak whereas, in the longer exposure images, the outer coronal signal is strong but the inner corona is over-exposed. Thus, the images need to be combined in such a way that the inner and outer corona are adequately exposed in the final image. This is done by using a signal-to-noise ratio (SNR) cut based on the square root of the intensity, thus providing a lower boundary wherein any pixels with a value less than this threshold are excluded. The SNR cut used in this work was $I > \sqrt{I}$ where $I$ is the intensity of the pixel. An upper threshold of 3,900 counts is also used since the linearity of the CMOS sensor breaks down around this value (see Figure \ref{fig:CIP linearity}). This process was done for each image taken at a given polarization angle which means that each composite image consists of two sets of different exposure times. \black

\red\subsubsection{Sky Brightness Removal}
\label{sec:sky_brightness_removal}

\begin{figure}[h!]
    \centering
    \includegraphics[width=0.85\textwidth]{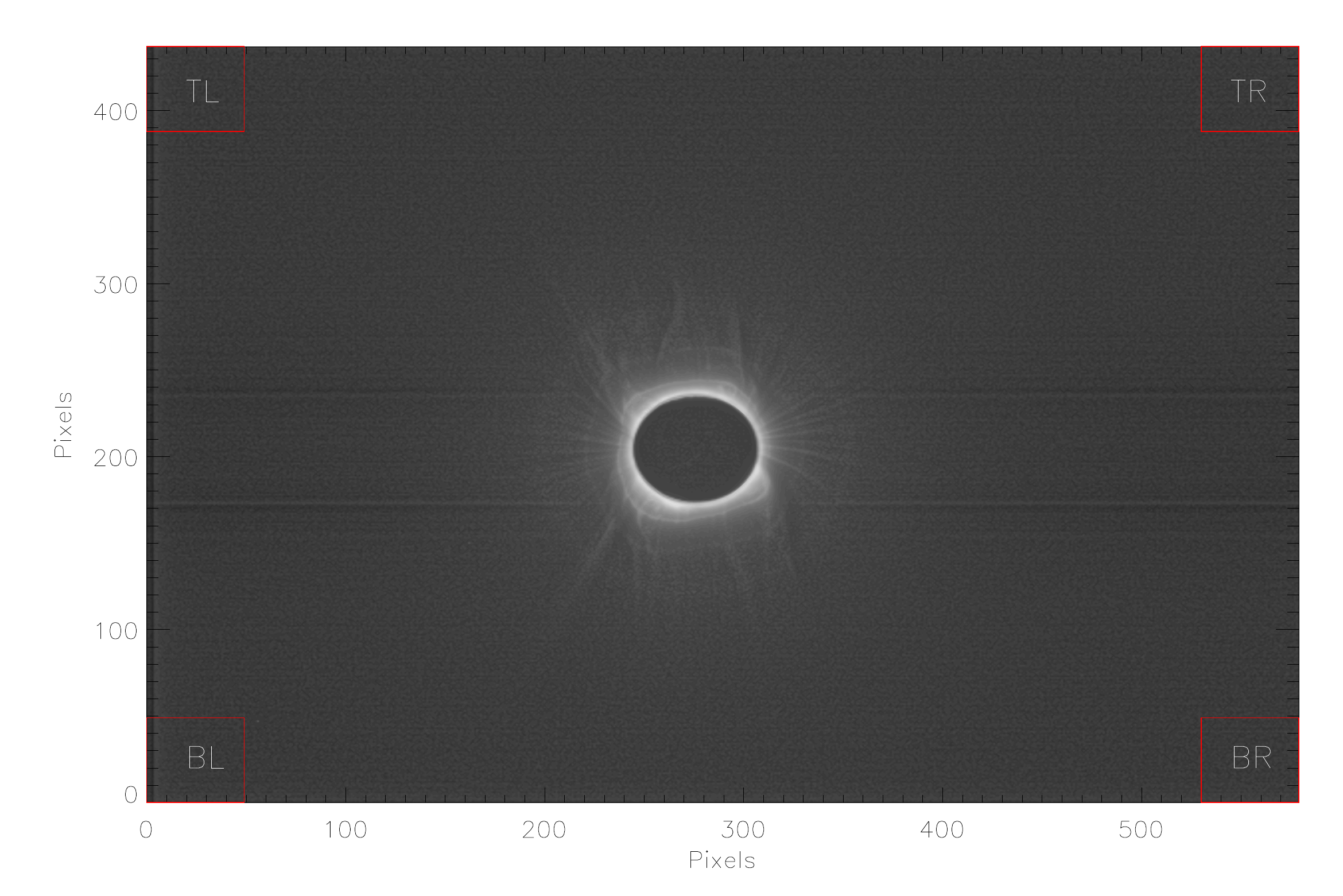}
    \caption{\red Location of each box used to determine the sky brightness overlaid on the combined 0\de \, image. The image has not been rotated so that solar north aligns with the top of the image. L, T, B, and R correspond to the left, top, bottom, and right sides of the image, respectively}
    \label{fig:boxes_positions_image}
\end{figure}

\begin{figure}[h!]
    \centering
    \includegraphics[width=0.85\textwidth]{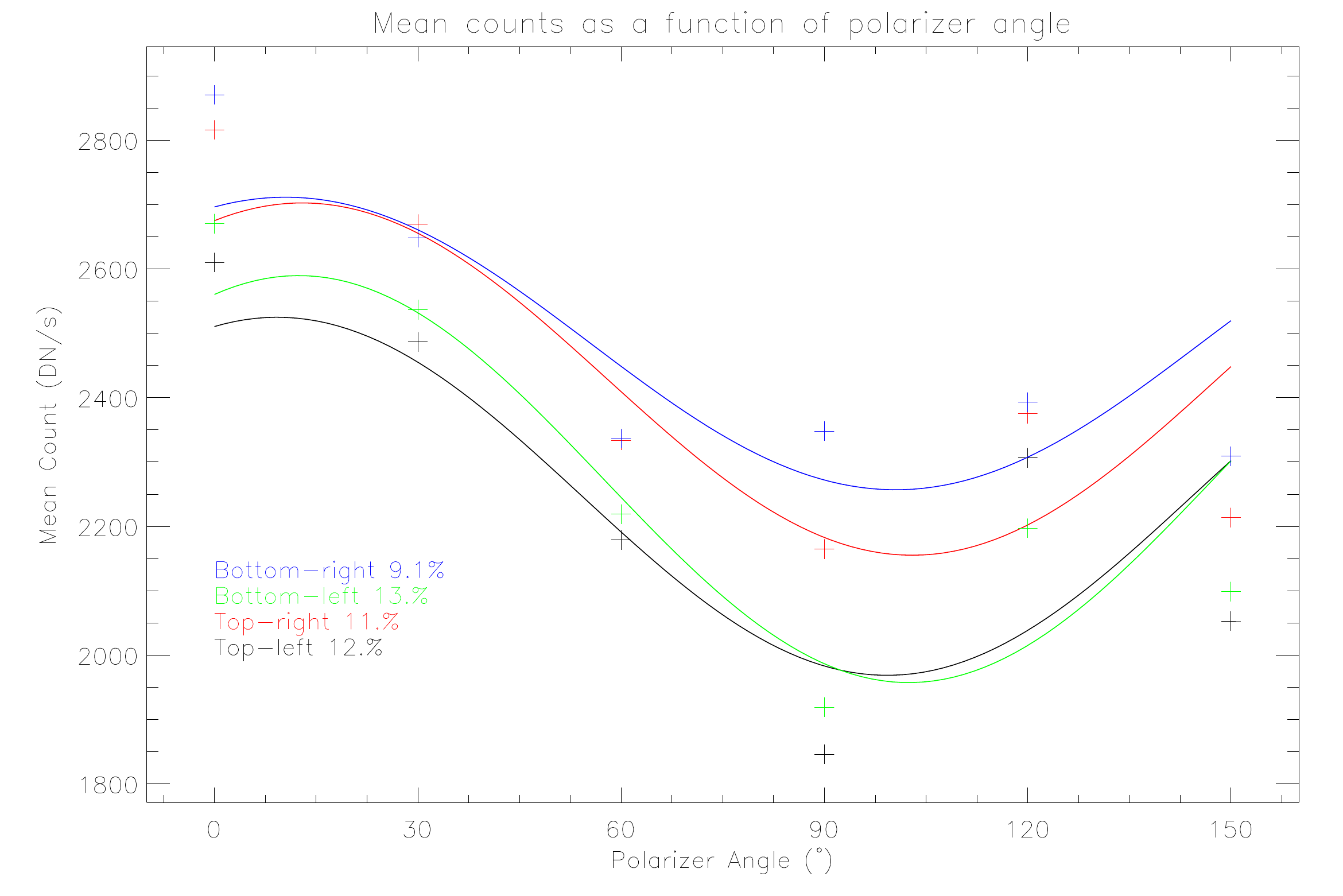}
    \caption{\red Mean intensity for each box in Figure \ref{fig:boxes_positions_image} at each polarizer angle. The fit to equation \ref{eq:measured_Intensity} is shown as a solid line for each of the four boxes where the results from the top-left, top-right, bottom-left, and bottom-right boxes are shown in black, red, green, and blue, respectively. The mean polarization of each box is also expressed as a percentage of the total polarization}
    \label{fig:mean_intensity_boxes_plot}
\end{figure}

\begin{figure}[h!]
    \centering
    \includegraphics[width=0.49\textwidth]{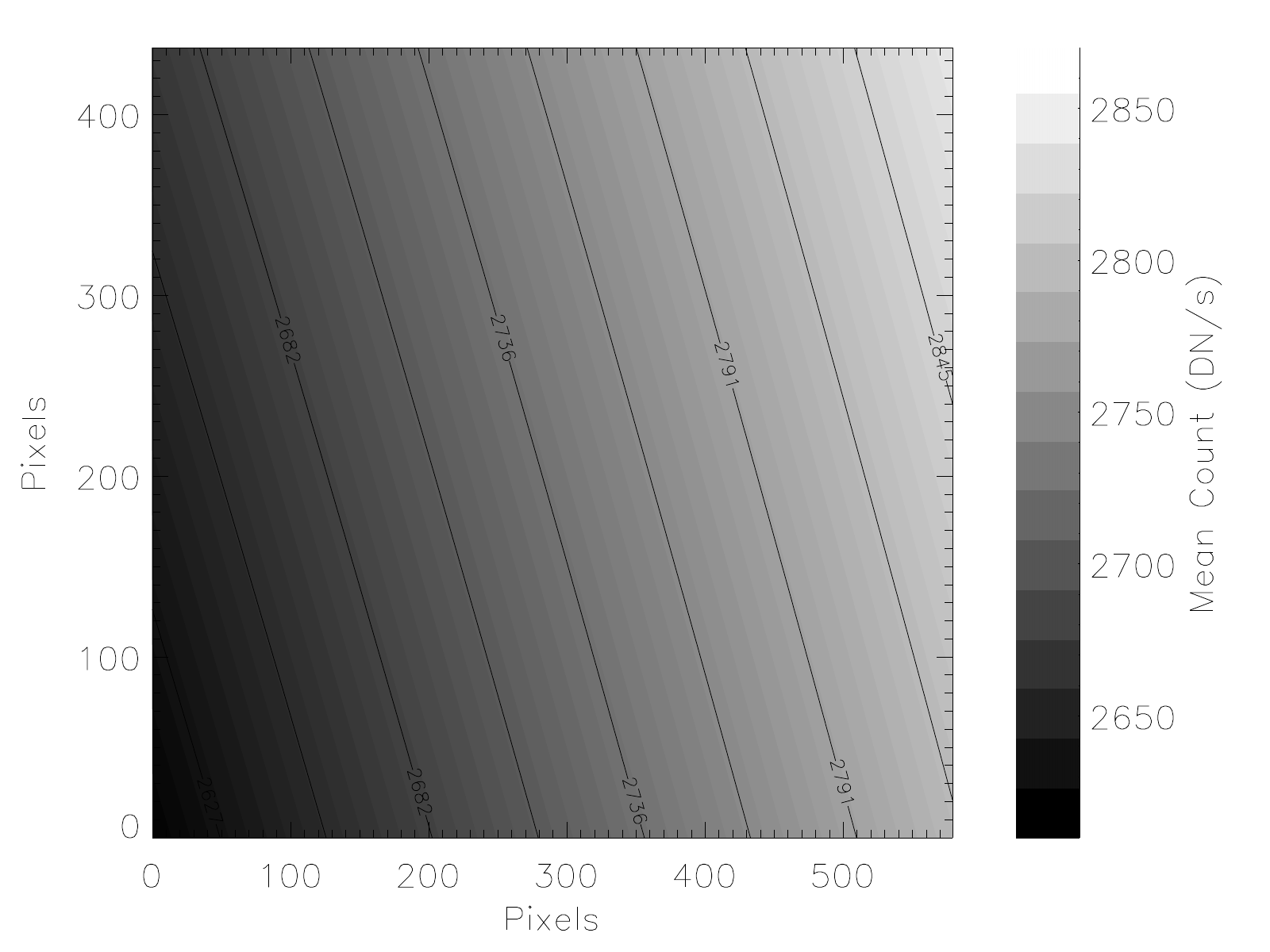}
    \hfill
    \includegraphics[width=0.49\textwidth]{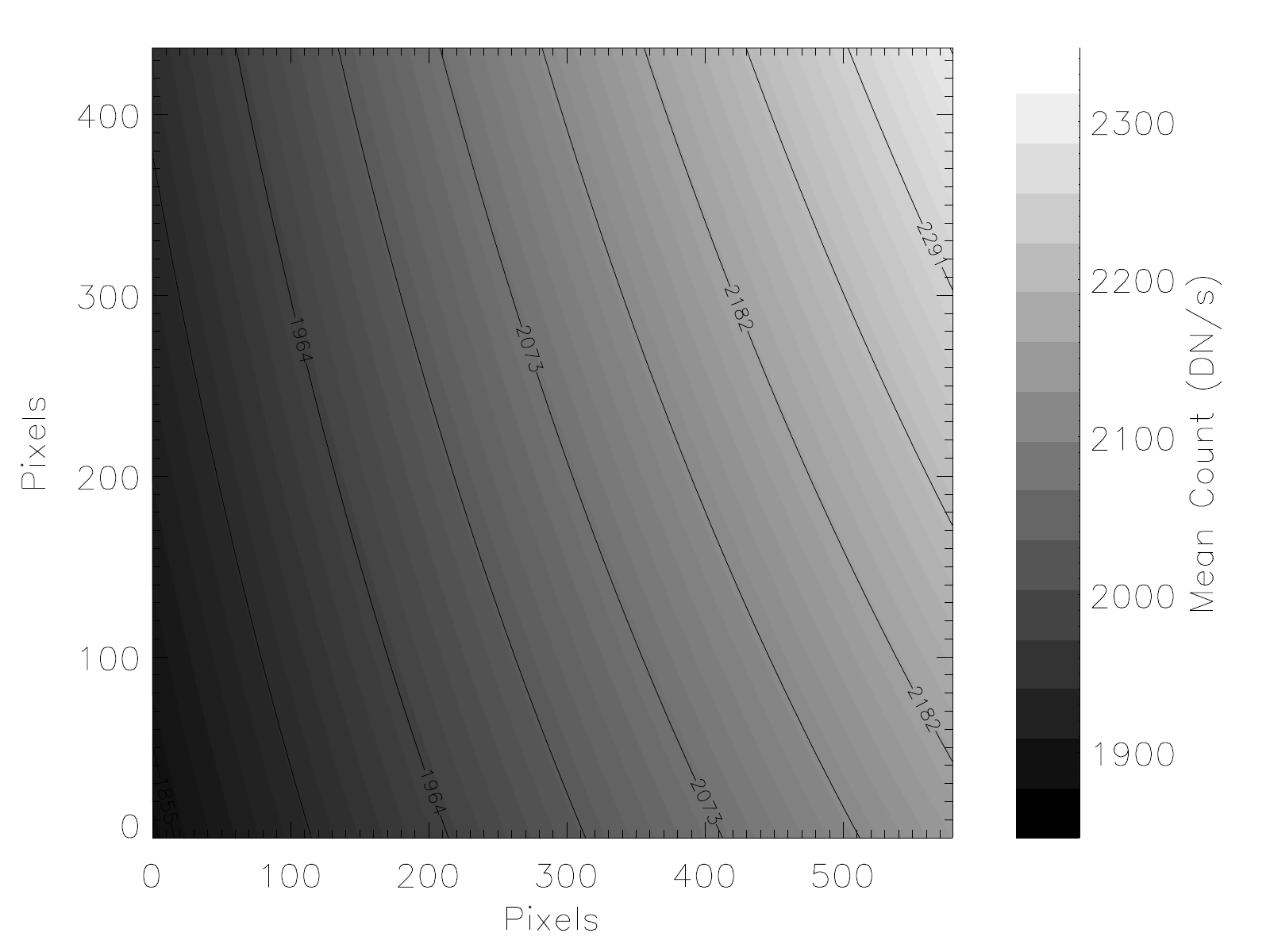}
    \caption{\red Examples of interpolated sky brightness maps subtracted from the respective TSE images for polarizer orientation angles of 0\de (\textit{left}) and 90\de (\textit{right})}
    \label{fig:interp_maps}
\end{figure}

During a TSE, the local sky brightness is significantly lowered to $\approx 10^{-9} - 10^{-10}$ B$_\odot$ which allows unmatched VL observations of the corona out to greater heliocentric heights than without a TSE. However, the sky brightness during a TSE is not negligible and must be removed from the data. As Figure \ref{fig:boxes_positions_image} shows, a box was defined in each corner of each combined TSE image. There is a considerable amount of noise present in the image as a result of the poor weather conditions, despite the rebinning to improve the signal-to-noise ratio. There is also an optical artifact that takes the form of two parallel lines spanning the width of the image, the cause of which is unknown but most probably instrumental. The mean intensity in each box was then calculated and the process was repeated for each polarizer angle. The result of this can be seen in Figure \ref{fig:mean_intensity_boxes_plot} which seems to show that the intensity at the right-hand side of the image (corresponding to solar north) is greater than that of the left-hand side (corresponding to solar south). A map is then created to estimate the sky brightness at each point in the image by interpolating the mean counts between each of the four boxes for each polarizer angle separately. Two examples of these interpolated sky brightness maps are shown in Figure \ref{fig:interp_maps} for 0\de \hspace{0.1mm} and 90\de. The polarization of the background sky lies at around 10\% of the overall polarization of the image and this component is subsequently removed as a result of this sky brightness removal step. Finally, the F-coronal component has a noticeable impact on $pB$ data from heliocentric heights of $\sim$ 2.5 - 3.0\Rs \, \citep{2021ApJ...912...44B} but since this study limited observations to a heliocentric height of 1.5\Rs, the F-coronal component can be neglected at such low heights.

\newpage

\black

\newpage
\subsubsection{Determining the Polarized Brightness} \label{subsec:polarization}
\vspace{2mm}

A standard inversion method used to acquire coronal electron densities was initially developed by \cite{1950BAN....11..135V} and further developed by \cite{1967ARA&A...5..213N}, \cite{1977SoPh...55..121S}, \cite{2001ApJ...548.1081H}, and \cite{2002A&A...393..295Q}, among others. The original model assumes both spherical symmetry and that the $pB$ is produced purely by the Thomson scattering of photospheric light from free coronal electrons and it is proportional to the integrated LOS density of the electrons, as shown in Equation (\ref{eq:density_inversion}). Most similar studies use three or four different polarization angles to determine the $pB$ - typically -60\de, 0\de, 60\de \, \citep{2021SoPh..296..158H}, and 0\de, 45\de, 90\de, 135\de \, \citep{2020PASP..132b4202V}. However, CIP \red captured images taken at different \black exposure times for six different polarization angles (0\de, 30\de, 60\de, 90\de, 120\de, and 150\de). As a result, in order to invert the calibrated intensities to find the $pB$ at each pixel, an approach involving least squares fitting was used. For a polarizer angle, $\theta_i$, where $i=0,1,2,...,n-1$ (with $n=6$ for \red this study\black), the measured intensity, $I_i$, is described by Equation \ref{eq:measured_Intensity}.

\begin{equation}
\label{eq:measured_Intensity}
I_i = I_0 + a \cos{2\theta_i} + b \red \sin \black {2\theta_i}.
\end{equation}

\red where the coefficients to be fitted \black are the unpolarized background intensity, $I_0$, and the polarized components, $a$ and $b$. The polarized brightness is then given by

\begin{equation}
    pB = \sqrt{a^2+b^2}
\end{equation}

\red In order to find a solution for $I_0$ and $pB$ at each pixel, the squared sum is then minimized as follows \black

\begin{equation}
\sum_{i=0}^{n-1} [I_i - (I_0 + a \cos{2\theta_i} + b \red\sin \black {2\theta_i})]^2
\end{equation}

\red This least squares fitting method was tested against the 
standard Mueller matrix inversion method for three polarization angles (0\de, 60\de, and 120\de) and it agreed within a few percent. \black As a result, CIP can theoretically take images of the corona for any number of polarizer angles in the future.

\red\subsubsection{Relative Radiometric Calibration}\black \label{subsec:kcor}
\vspace{2mm}

\begin{figure}[h!]
    \centering
    \includegraphics[width=0.68\textwidth]{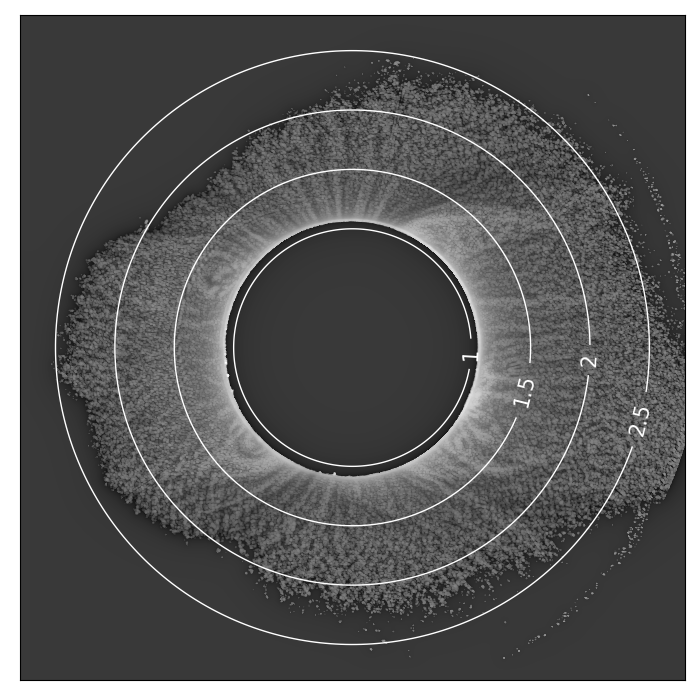}
    \caption{\red 10-minute averaged $pB$ observations taken by MLSO's K-Cor instrument on the day of the eclipse, processed using the Multi-scale Gaussian Normalization technique, with the concentric circles corresponding to heliocentric heights of 1.0, 1.5, 2.0, and 2.5\Rs, respectively\black}
    \label{fig:k-cor cartesian}
\end{figure}

A relative radiometric calibration step converts the polarized brightness measured by CIP (recorded as DN/s) to units relative to the mean solar brightness (MSB)\red. The instrument used in this step was the Mauna Loa Solar Observatory's (MLSO) COSMO K-Coronagraph (DOI: \href{https://mlso.hao.ucar.edu/mlso_data_calendar.php}{10.5065/D69G5JV8}) which provides $pB$ data with a field-of-view from 1.05 to \red $\sim$ 3\Rs \, and a spatial resolution of 11.3". The 10-minute averaged K-Cor data was used for this calibration step with the first observation occurring at 17:56:41 UTC and the last at 18:10:20 UTC. Figure \ref{fig:k-cor cartesian} shows the 10-minute averaged $pB$ observations taken by MLSO's K-Cor instrument on the day of the TSE, processed using the Multi-scale Gaussian Normalization technique \citep{2014SoPh..289.2945M} in order to enhance the fine-scale coronal structure, and the concentric rings represent heliocentric heights of 1.0, 1.5, 2.0, and 2.5\Rs, respectively. It is clear to see that consistent $pB$ data is restricted to $\sim$ 1.7\Rs with the data extending further out in the equatorial regions. As a result, the relative radiometric calibration in this study was limited to a heliocentric height of 1.5\Rs. \black

\begin{figure}[h!]
    \centering
    \includegraphics[width=0.49\textwidth]{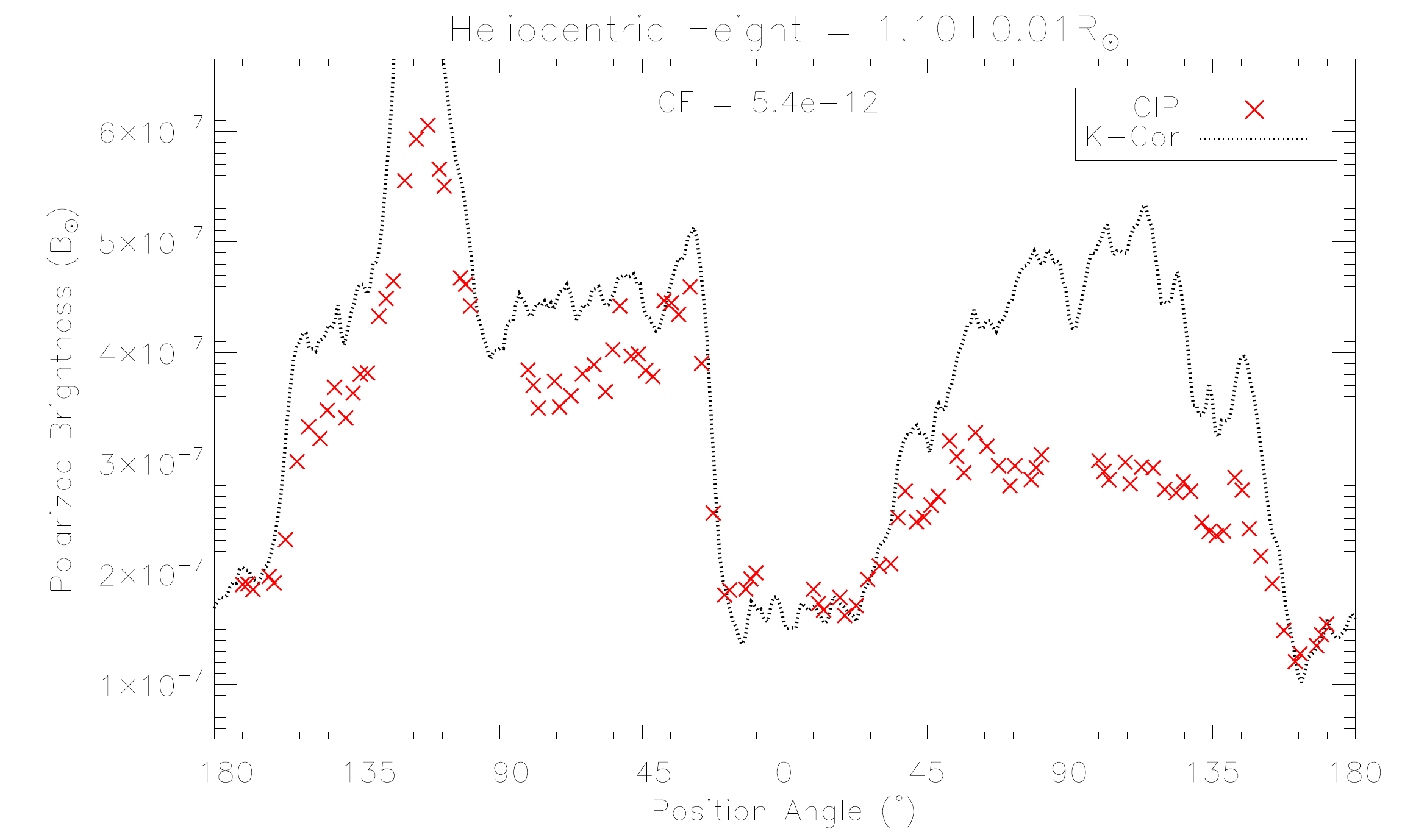}
    \hfill
    \includegraphics[width=0.49\textwidth]{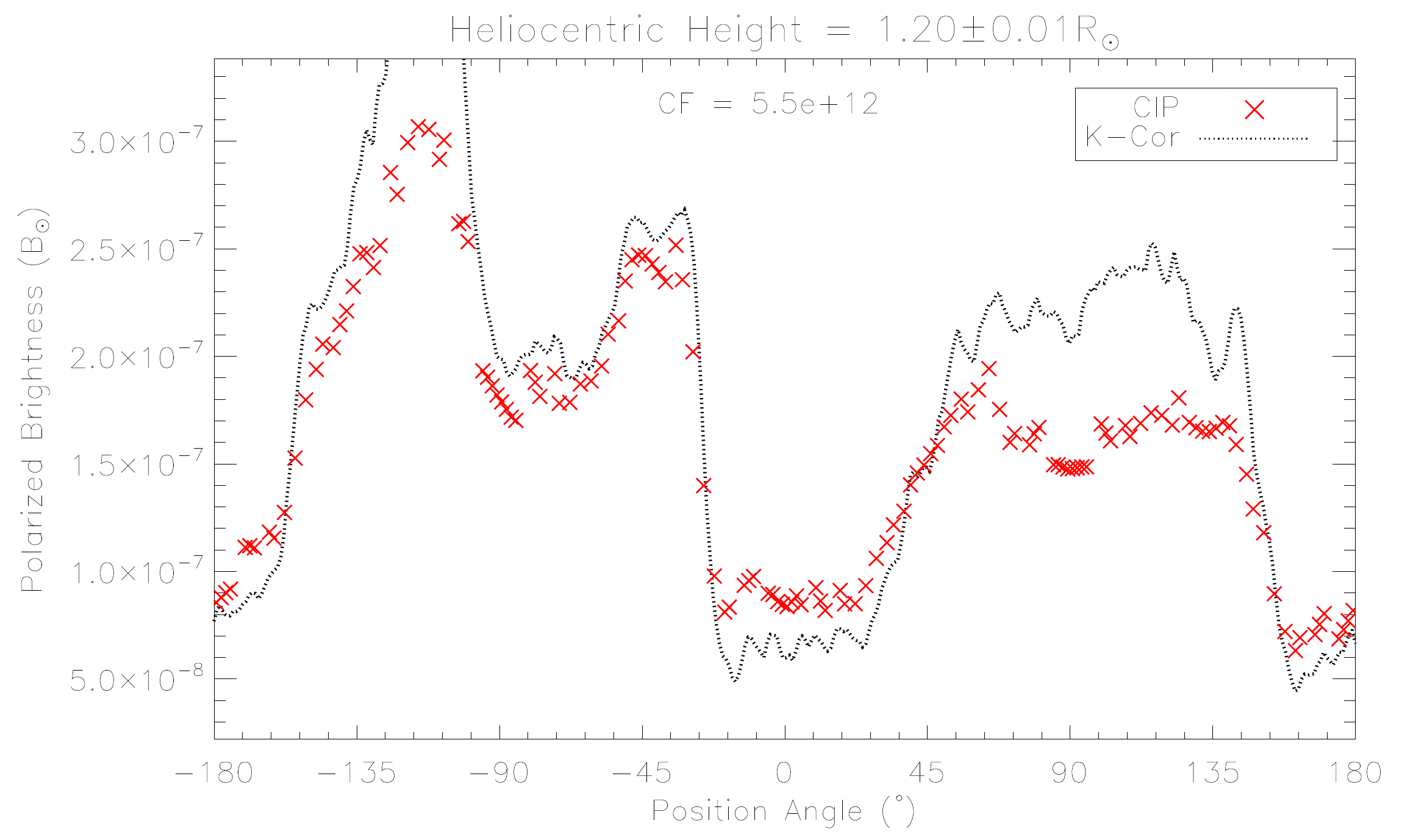}
    \vfill
    \includegraphics[width=0.49\textwidth]{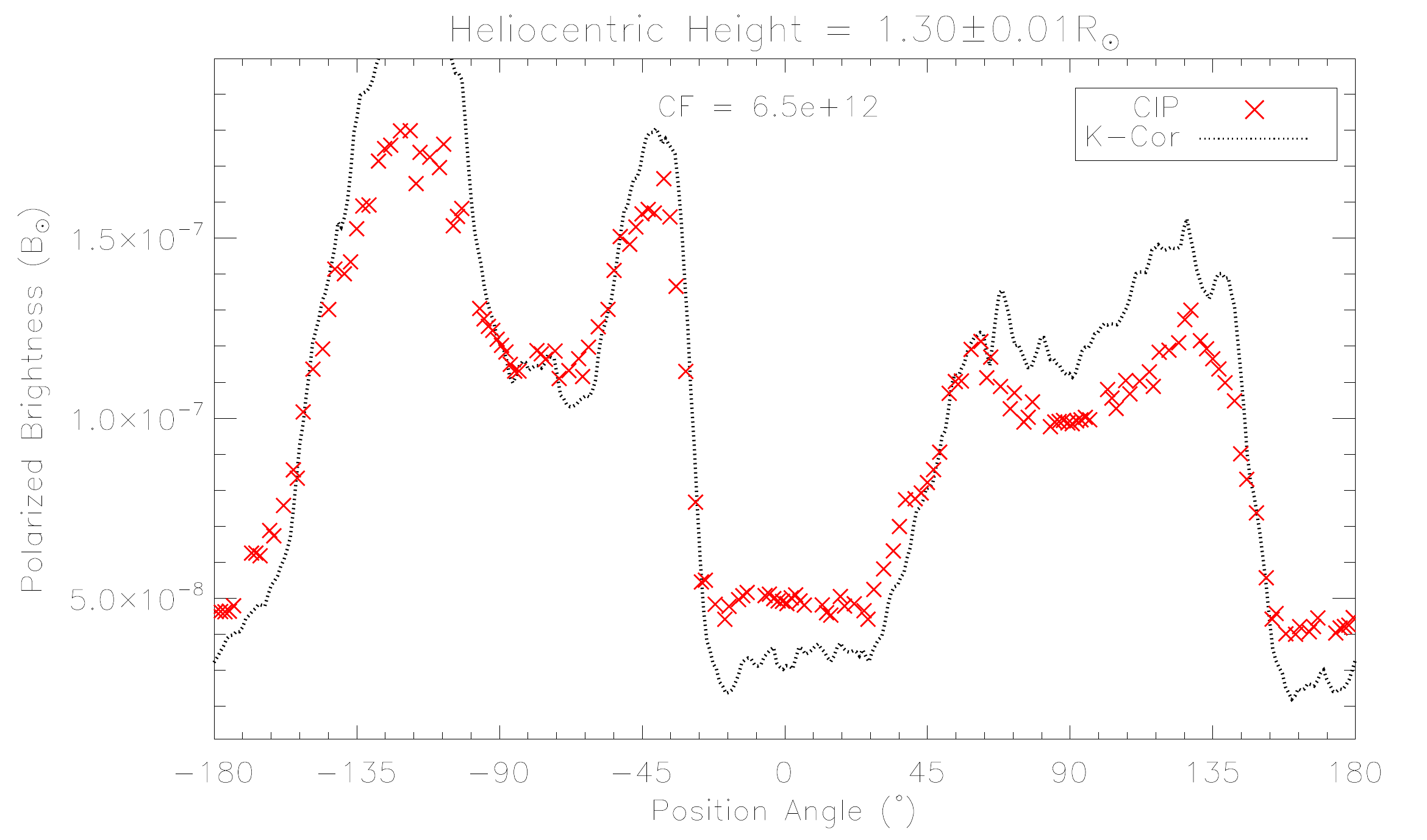}
    \hfill
    \includegraphics[width=0.49\textwidth]{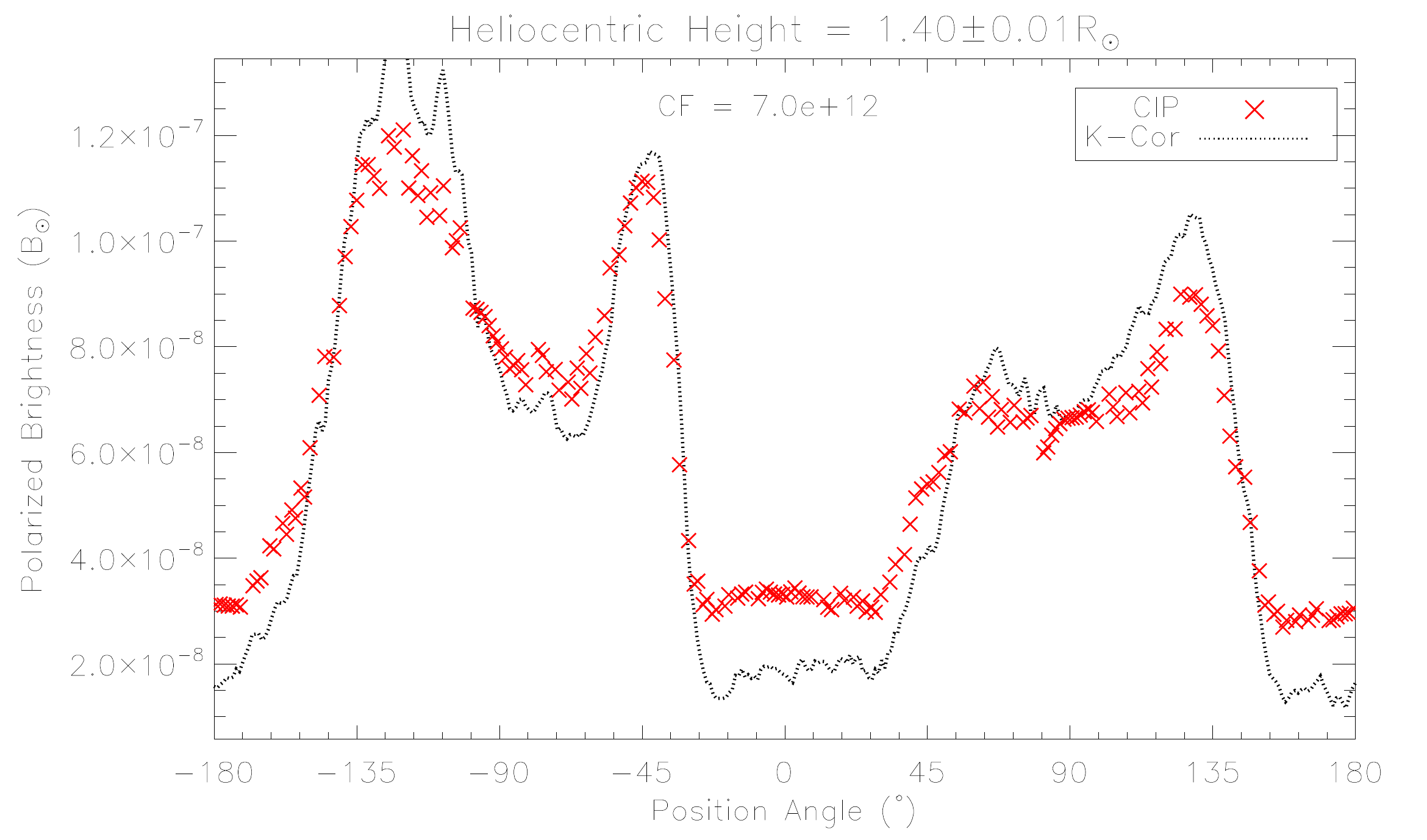}
    \vfill
    \includegraphics[width=0.49\textwidth]{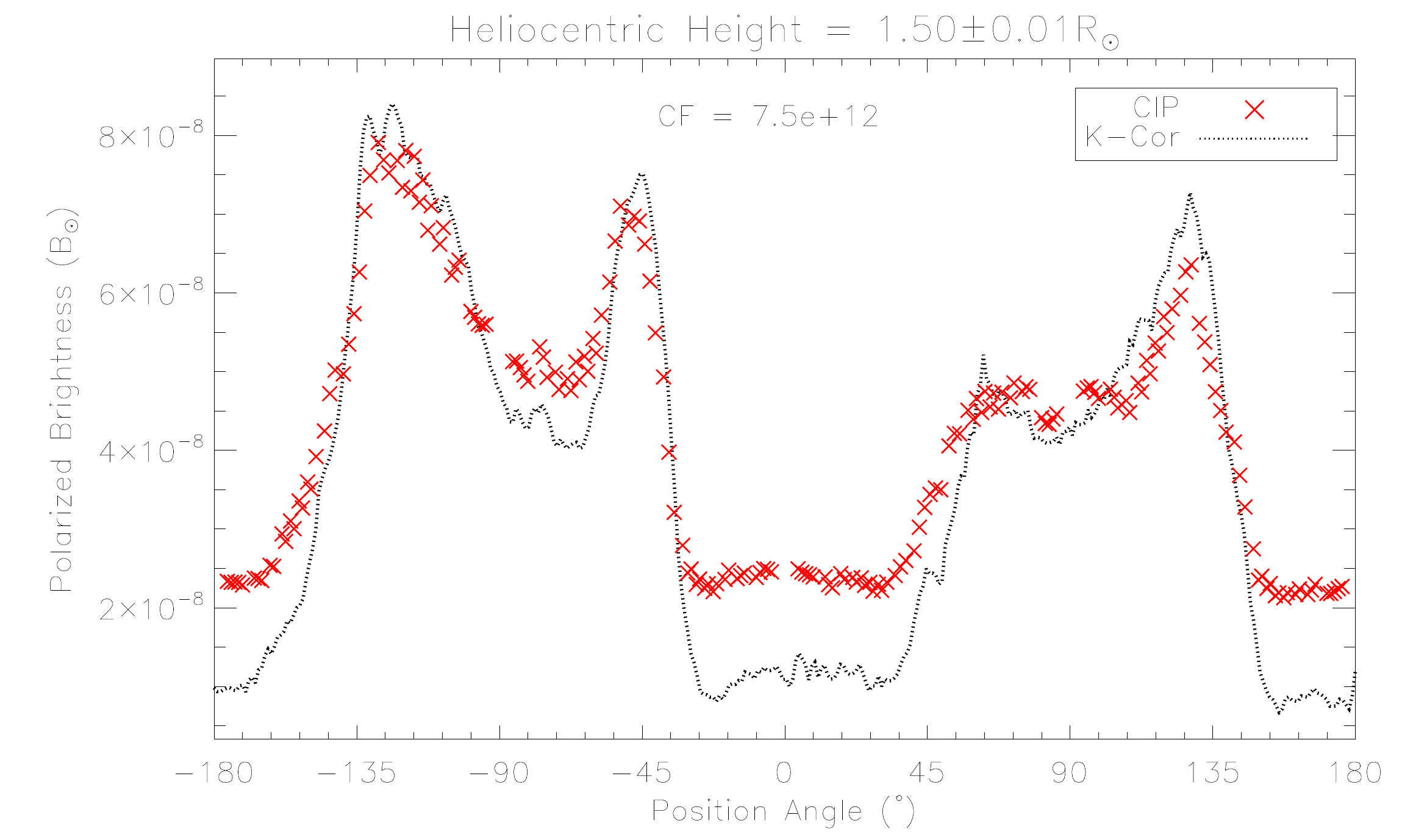}
    \caption{\red Latitudinal distribution of intensity profiles from both CIP (red) and K-Cor (black) for a range of heliocentric heights. A calibration factor (CF), noted in each plot, is temporarily applied to the CIP data in order to see both profiles on the same axis scale\black}
    \label{fig:pb_lat_dist}
\end{figure}

\begin{figure}[h!]
    \centering
    \includegraphics[width=0.8\textwidth]{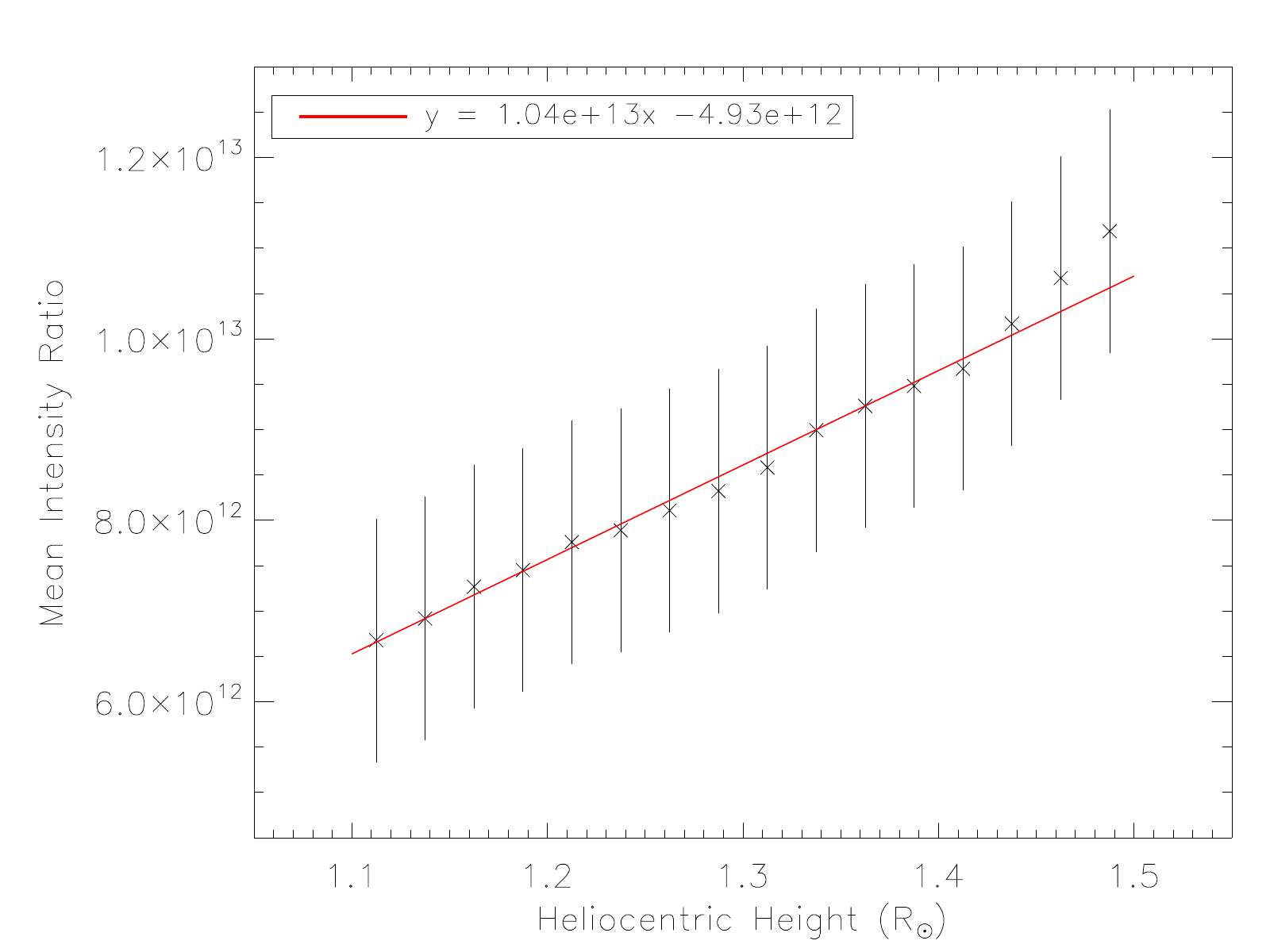}
    \caption{\red Mean intensity ratio between the CIP and K-Cor intensity profiles for each height interval (black) along with the linear fit used to perform the relative radiometric calibration of the TSE data (red)}
    \label{fig:mean_ratios}
\end{figure}

\red The images from both CIP and K-Cor are then coaligned and an intensity profile is taken for a thin slice of the corona at a specified height for both images. Some examples of these intensity profiles can be seen in Figure \ref{fig:pb_lat_dist} for a range of heliocentric heights. A calibration factor is temporarily applied to the CIP data in order to visualize both latitudinal profiles on the same axis scale, which is denoted in each plot as CF. The mean ratio of the data from both instruments is then computed at intervals of 0.025\Rs between 1.1 and 1.5\Rs and these values are shown in Figure \ref{fig:mean_ratios}. This linear increase in the intensity ratio with height is then applied to the TSE data thus converting the $pB$ observations taken with CIP from units of DN/s to units of solar brightness (\Bs). A cross-correlation was also performed on the intensity profiles to find the angle needed to rotate the CIP data in order to align it with that taken by K-Cor (i.e., solar north upwards). \black
\black

\newpage

\section{Results}
\label{sec:results}


\begin{figure}[h!]
    \centering
    \includegraphics[width=0.6\textwidth]{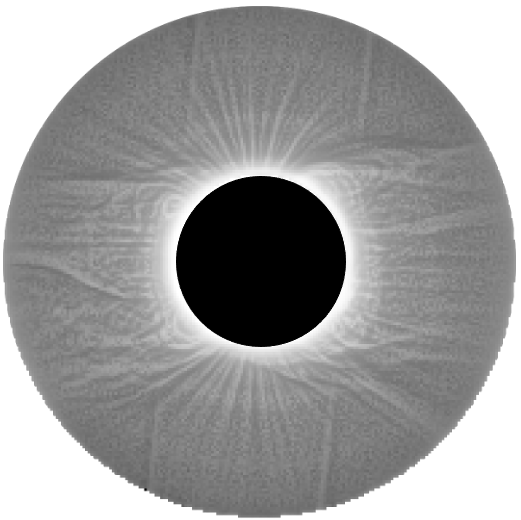}
    \caption{\red Final MGN-processed $pB$ image of the December 14 2020 TSE after relative radiometric calibration with respect to MLSO/K-Cor}
    \label{fig:final_calibrated_pb_image}
\end{figure}

\newpage

\red Figure \ref{fig:final_calibrated_pb_image} shows the final calibrated $pB$ image of the December 14 2020 TSE. It has been further processed using the MGN technique to enhance the fine-scale coronal detail. Due to the poor viewing conditions at the observation site adversely affecting the data, the data analysis has been limited to a heliocentric height of 1.5\Rs but the image in Figure \ref{fig:final_calibrated_pb_image} has been extended out to 2\Rs to show more of the corona.

\begin{figure}[h!]
    \centering
    \includegraphics[width=0.58\textwidth]{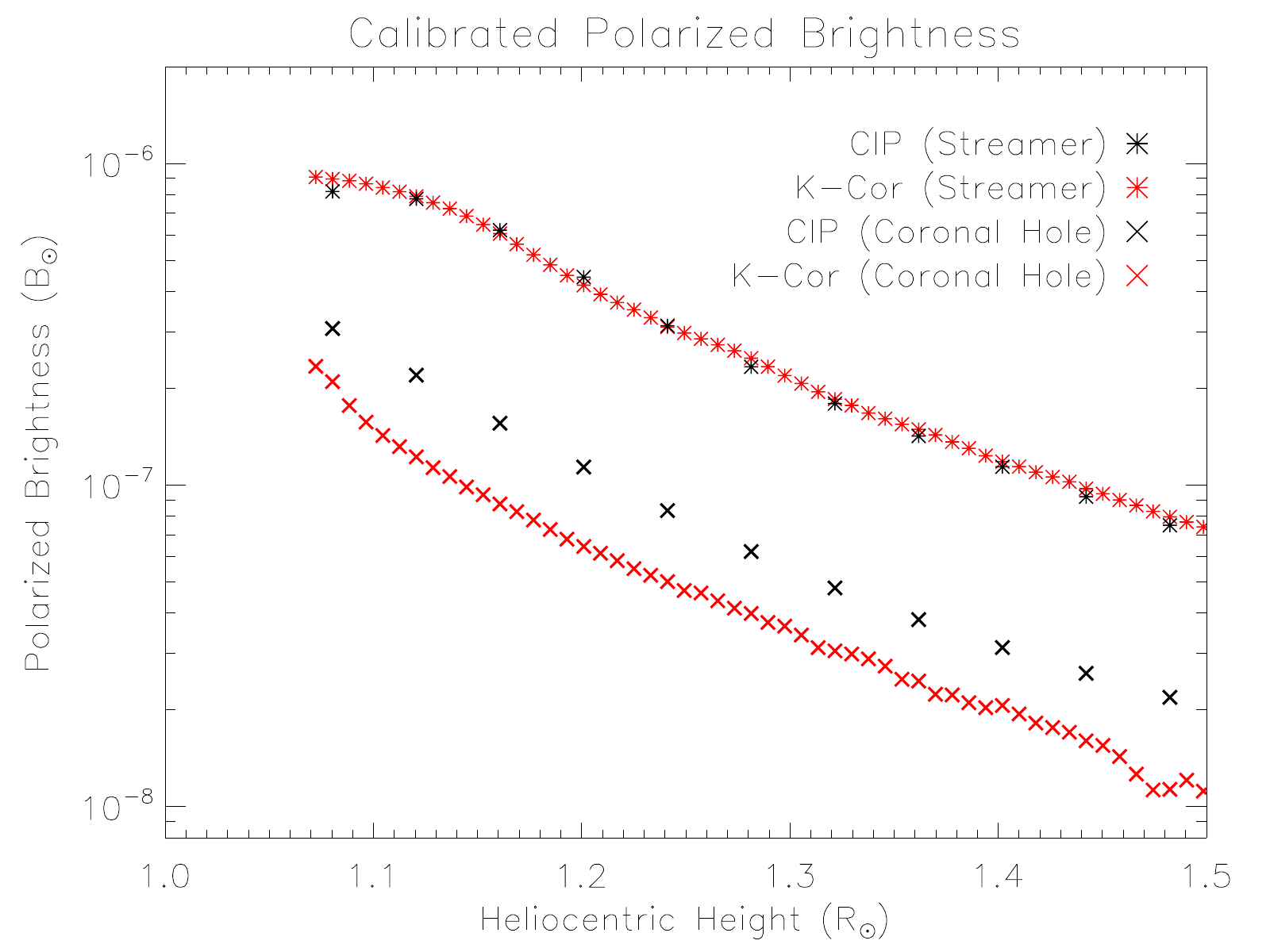}
    \caption{\red Calibrated and corrected $pB$ profiles from CIP (black) and K-Cor (red) for both the polar coronal hole and equatorial streamer}
    \label{fig:calibrated_pb_data}
\end{figure}

\red Figure \ref{fig:calibrated_pb_data} shows a comparison of $pB$ data as a function of heliocentric height for the north polar coronal hole and west equatorial streamer chosen for this study (at position angles 12\de \hspace{0.1mm} and 245\de \hspace{0.1mm} counter-clockwise from solar north, respectively). As would be expected, the $pB$ is highest in the solar equatorial region and lower in the polar region. It is clear to see that the relative radiometric calibration with respect to MLSO/K-Cor is optimal at the equator but the polar $pB$ observed by CIP is still greater than that of K-Cor. This is believed to be due to the poor weather conditions at the time of observation having a greater effect on the fainter coronal signal in the polar regions.

\begin{figure}[h!]
    \centering
    \includegraphics[width=0.58\textwidth]{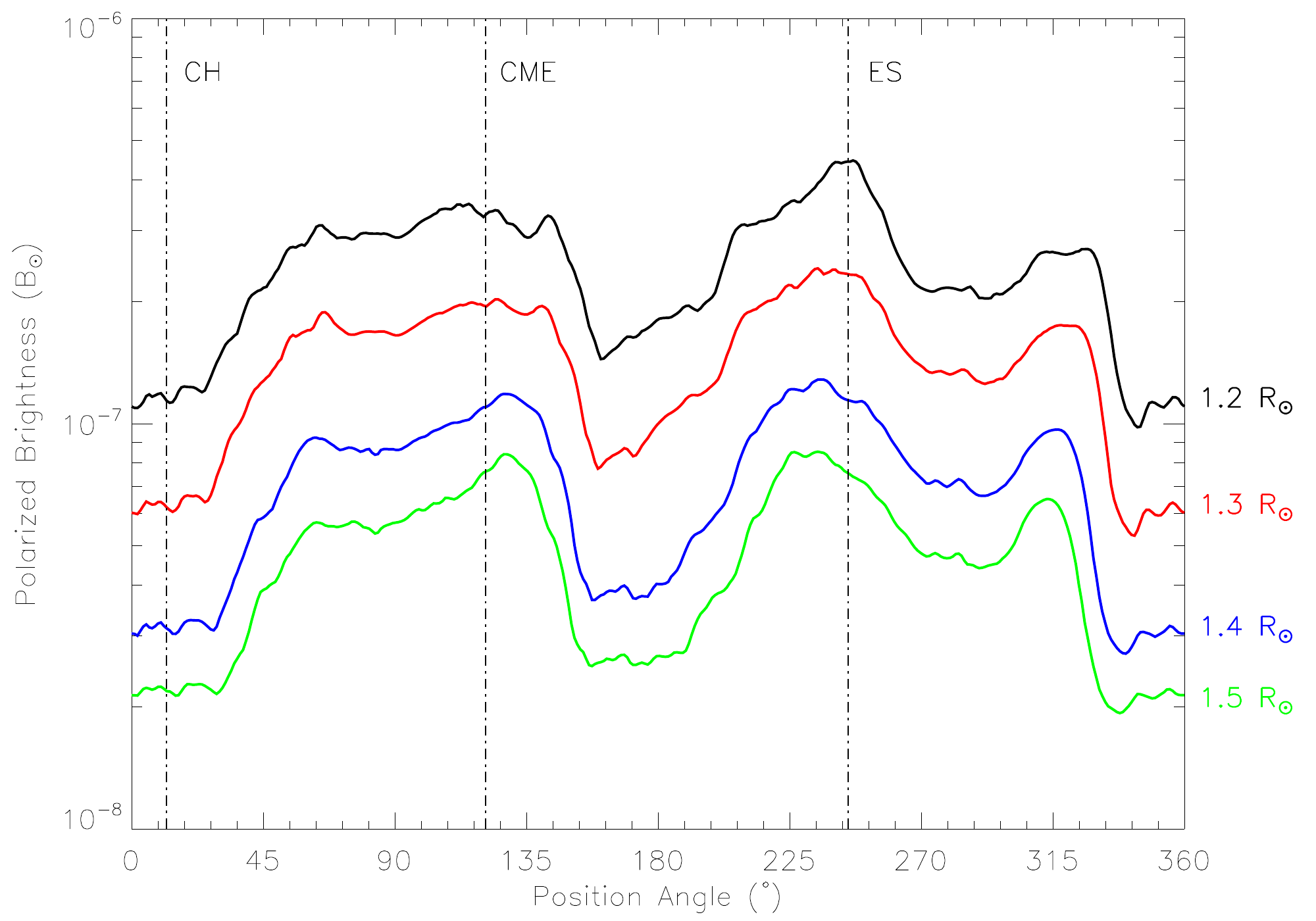}
    \caption{\red Latitudinal distribution of $pB$ for heliocentric heights of 1.2 (black), 1.3 (red), 1.4 (blue), and 1.5 \Rs (green). Vertical dotted lines are included to represent the position angles for the polar coronal hole (left), CME (middle), and equatorial streamer (right).}
    \label{fig:pb-latitudinal-distribution}
\end{figure}

\newpage

Figure \ref{fig:pb-latitudinal-distribution} shows the latitudinal distribution of the $pB$ for heliocentric heights of 1.2, 1.3, 1.4, and 1.5\Rs \hspace{0.1mm} in black, red, blue, and green, respectively. The $pB$ decreases with increasing distance from the limb which is to be expected, and all four latitudinal distributions have a similar shape but there is a noticeable offset between corresponding maxima and minima as heliocentric height increases. The dotted vertical lines represent the position angles at which the densities are calculated for the polar coronal hole and the equatorial streamer, as well as the position angle for the CME where its presence becomes increasingly apparent in the 1.4 and 1.5\Rs \, $pB$ distributions.

\begin{figure}[h!]
    \centering
    \includegraphics[width=0.49\textwidth]{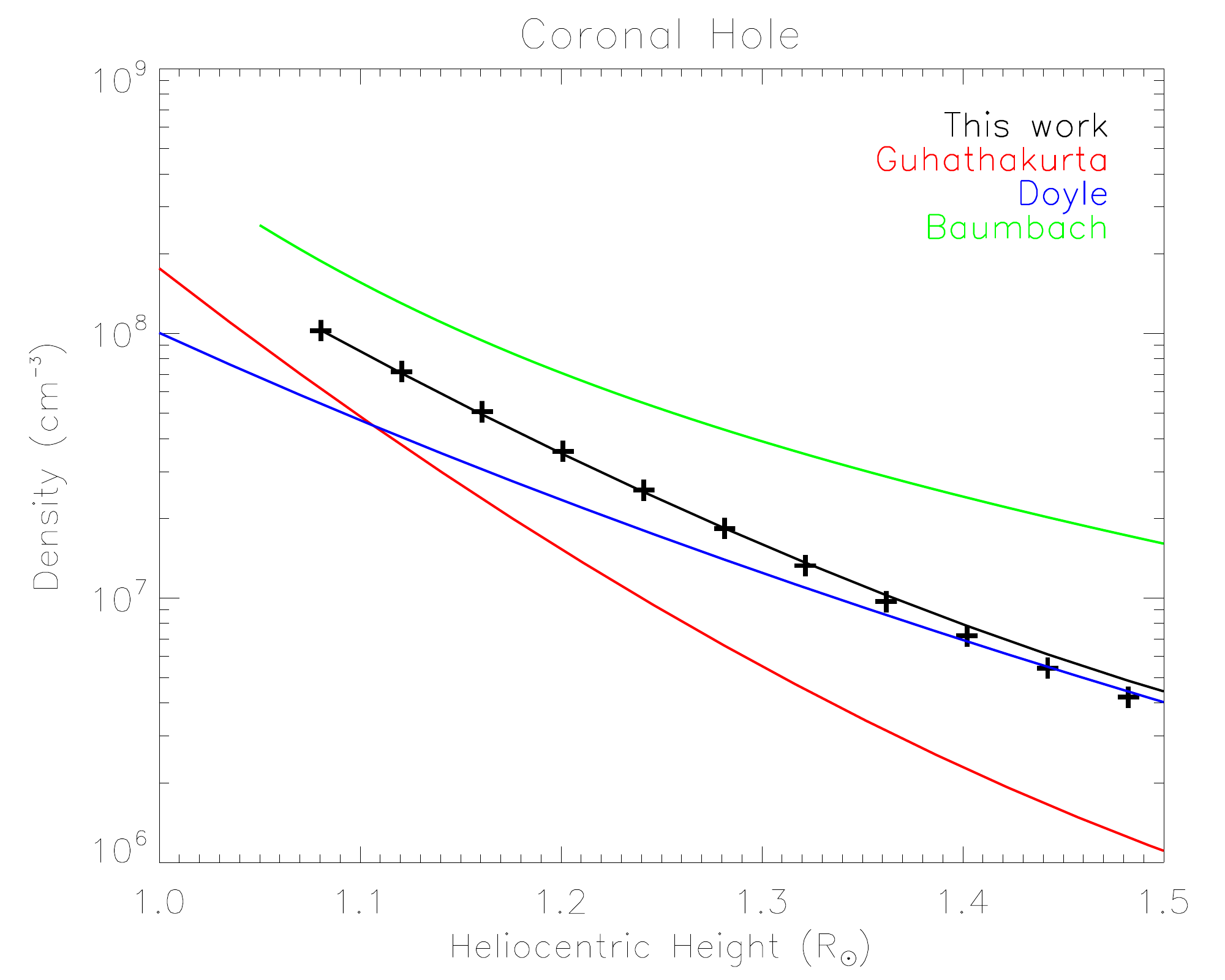}
    \hfill
    \includegraphics[width=0.49\textwidth]{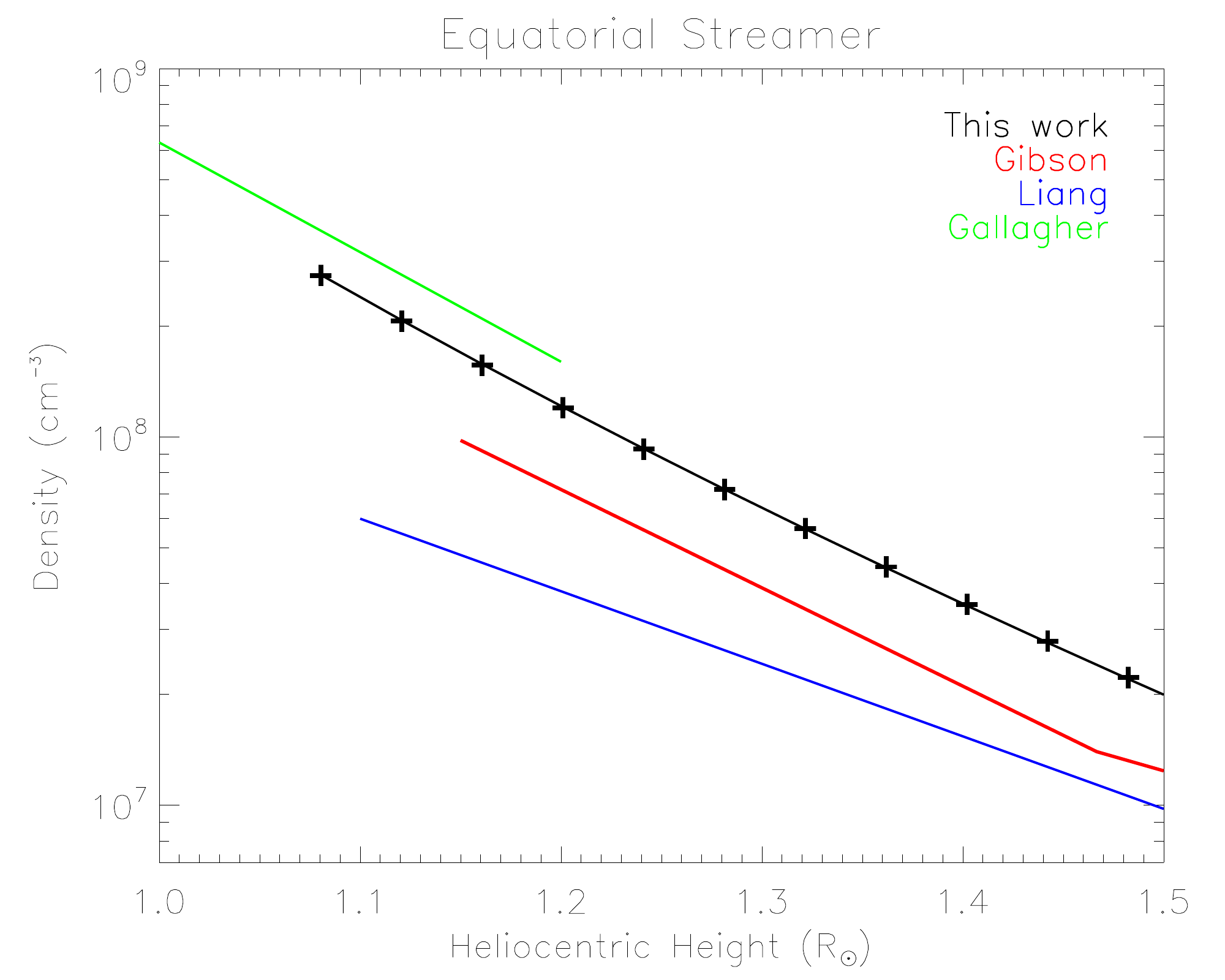}
    \caption{\red Coronal electron densities as a function of heliocentric height for the coronal hole (left) compared with \cite{baumbach}, \cite{1999A&A...349..956D}, and \cite{1999JGR...104.9801G}, and the equatorial streamer (right) compared with \cite{1999JGR...104.9691G}, \cite{2022MNRAS.tmp.2971L}, and \cite{1999ApJ...524L.133G}. The black data points show the densities acquired by inverting the $pB$ and the solid black line shows the fit to equation \ref{eq:coefficients}.}
    \label{fig:densities}
\end{figure}

Figure \ref{fig:densities} shows the coronal electron densities, derived by inverting the $pB$, as a function of heliocentric height for both the polar coronal hole (left) and equatorial streamer (right). Both plots also show comparisons to data from previous works which agree very well with the data from this study. As expected, the streamer's density is greater than the coronal hole's by an average factor of $\sim$ 4 between 1.1 - 1.5\Rs. For the coronal hole comparison, density profiles from \cite{baumbach}, \cite{1999A&A...349..956D}, and \cite{1999JGR...104.9801G} were used. The latter two in particular were selected because they represent observations of polar coronal holes taken at the start of solar cycle 24 (solar minimum) and, since the 2020 TSE occurred exactly a year after the start of solar cycle 25, they are reasonable comparisons to make. For the equatorial streamer, the densities obtained in this study are compared with \cite{1999JGR...104.9691G}, \cite{2022MNRAS.tmp.2971L}, and \cite{1999ApJ...524L.133G}. Again, these are reasonable comparisons to make since the compared works represent observations of a streamer (\cite{1999JGR...104.9691G}) and equatorial regions (\cite{2022MNRAS.tmp.2971L, 1999ApJ...524L.133G}) taken at or near solar minimum.

\black
\newpage

\section{Discussion}
\label{sec:discussion}

\red
\subsection{Number of Polarizer Angles}

\begin{figure}[h!]
    \centering
    \includegraphics[width=0.7\textwidth]{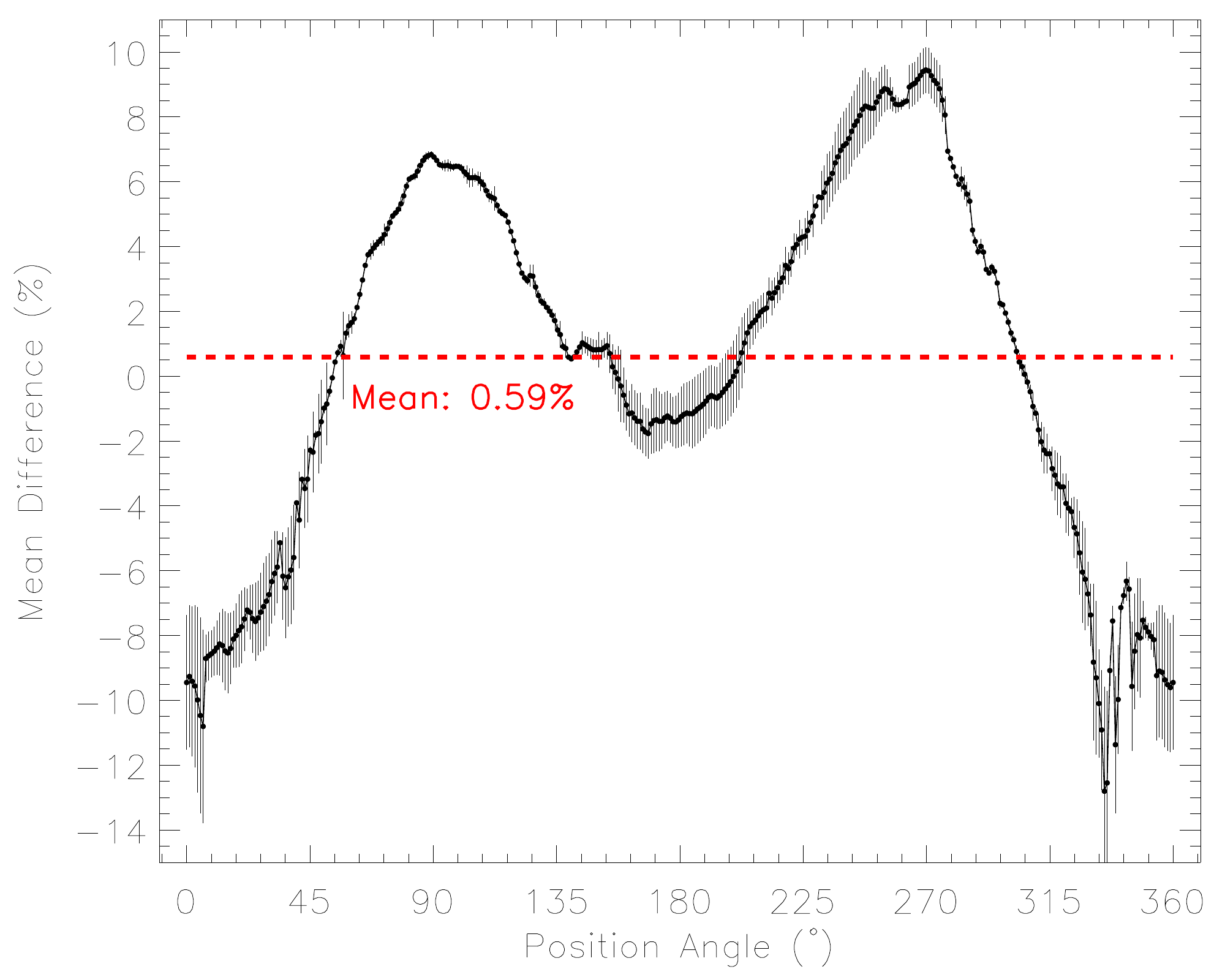}
    \caption{\red Mean difference between the coronal electron densities derived using all six polarizer angles (Angle Set A) and only the 0\de, 60\de, and 120\de polarizer angles (Angle Set B). The mean difference ranges from $\sim$ -13\% to 10\% and the mean in the difference is also shown (red line). The error bars show the standard error in the mean.}
    \label{fig:mean_diff_plot}
\end{figure}

Typically, similar studies and space-based coronagraphs use polarizer angles of 0\de, 60\de, and 120\de \, to infer the coronal electron density. As mentioned previously, CIP is designed to efficiently take polarized observations for any number of pre-defined polarizer angles. For this study, six polarizer angles were chosen in an attempt to see if increasing the number of polarizer angles leads to better constraints on the $pB$ observations and the inferred coronal electron densities. Figure \ref{fig:mean_diff_plot} shows the mean percentage difference between the coronal electron densities derived using all six polarizer angles (henceforth referred to as Angle Set A) and only the 0\de, 60\de, and 120\de polarizer angles (Angle Set B) with the error bars representing the standard error in the mean. The difference is shown as a function of the position angle between the heliocentric heights 1.1 - 1.5\Rs \hspace{0.1mm} and ranges from $-12.8\%$ to 9.44\% with the overall mean at 0.56\%. For the position angles chosen to represent the coronal hole and equatorial streamer, the mean difference is -8.26\% and 8.04\%, respectively.

\begin{figure}[h!]
    \centering
    \includegraphics[width=0.49\textwidth]{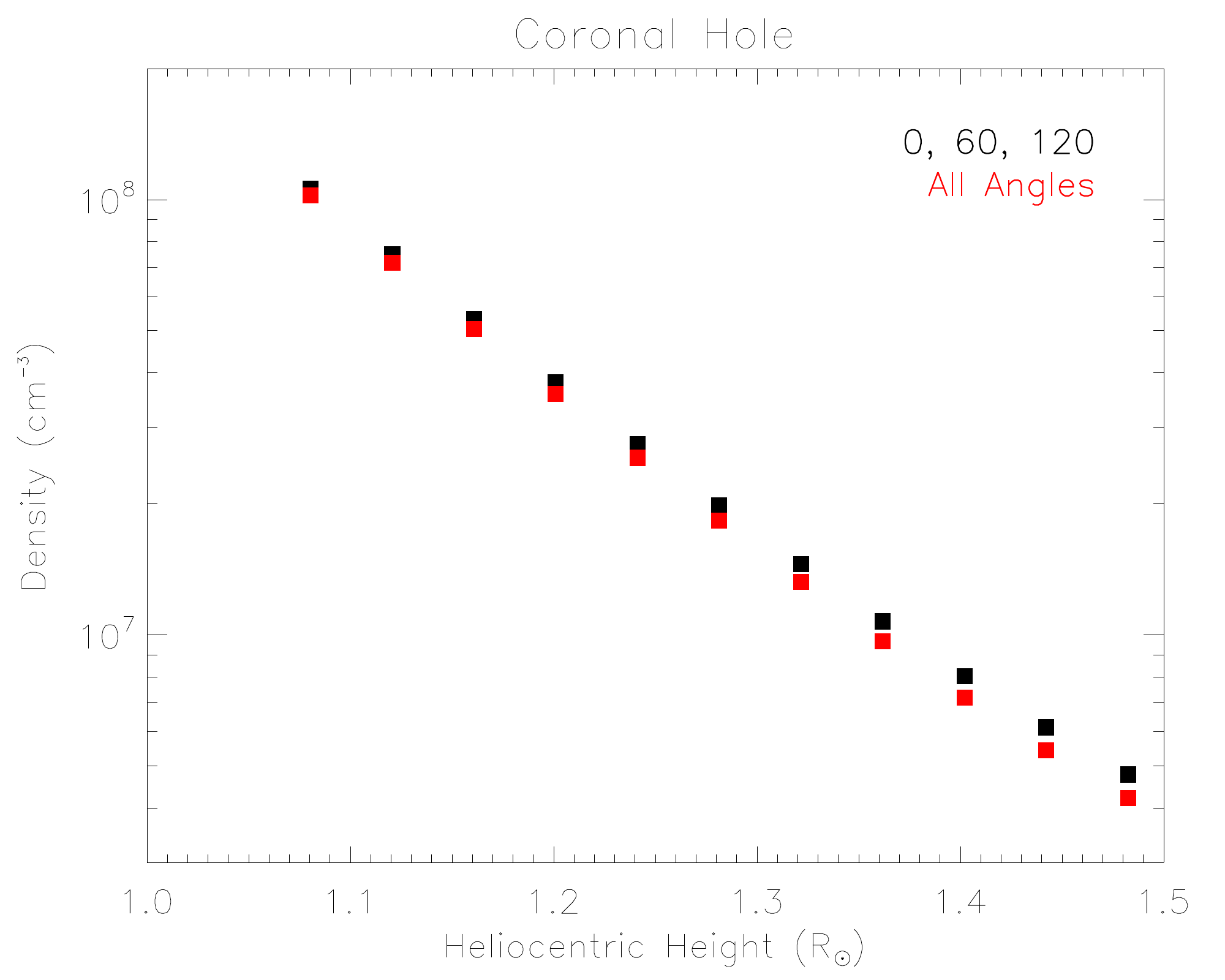}
    \hfill
    \includegraphics[width=0.49\textwidth]{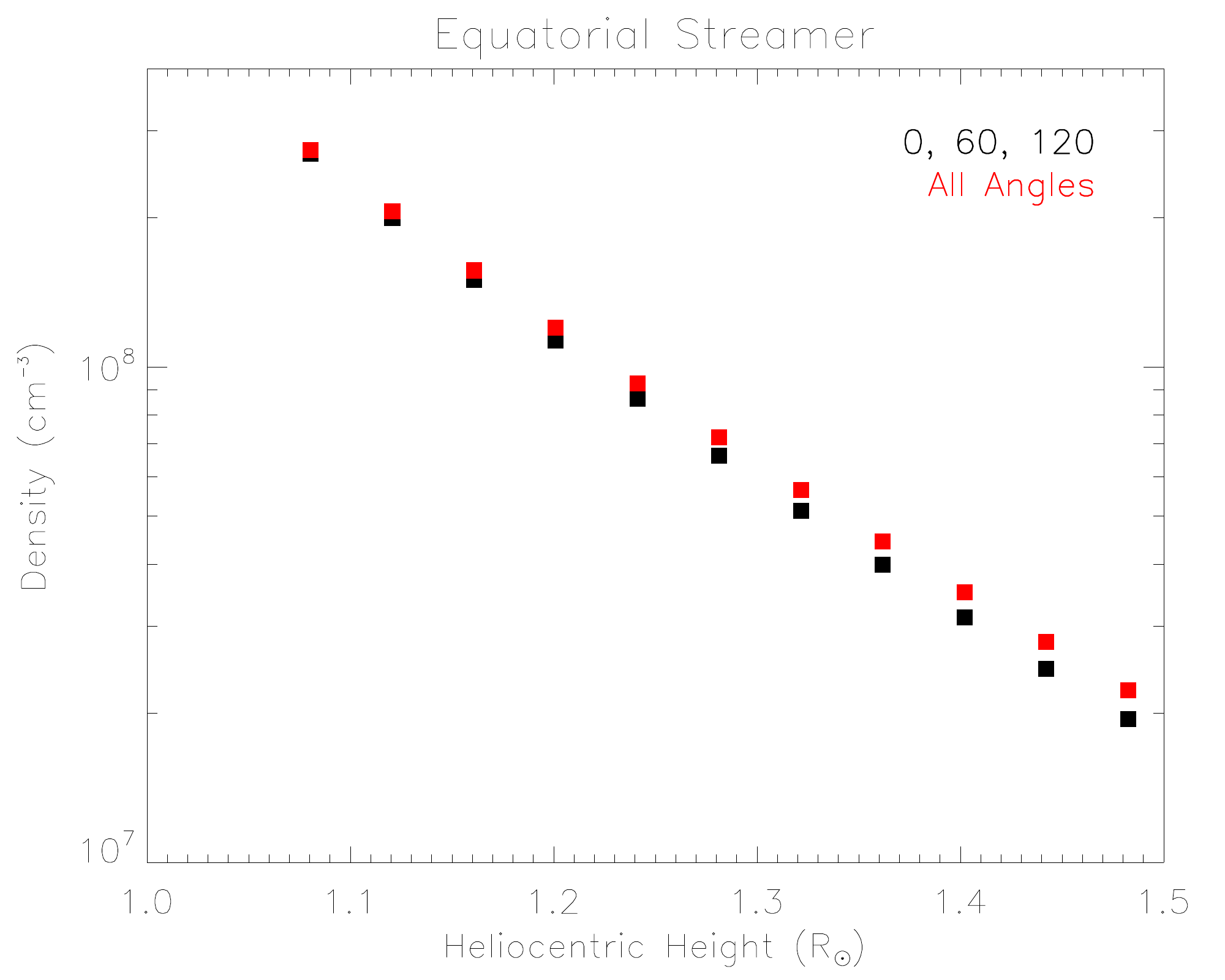}
    \caption{\red Difference in coronal electron densities between using $pB$ data from Angle Sets A (red) and B (black) for both the coronal hole and equatorial streamer}
    \label{fig:ne_comparison}
\end{figure}

\begin{figure}[h!]
    \centering
    \includegraphics[width=0.75\textwidth]{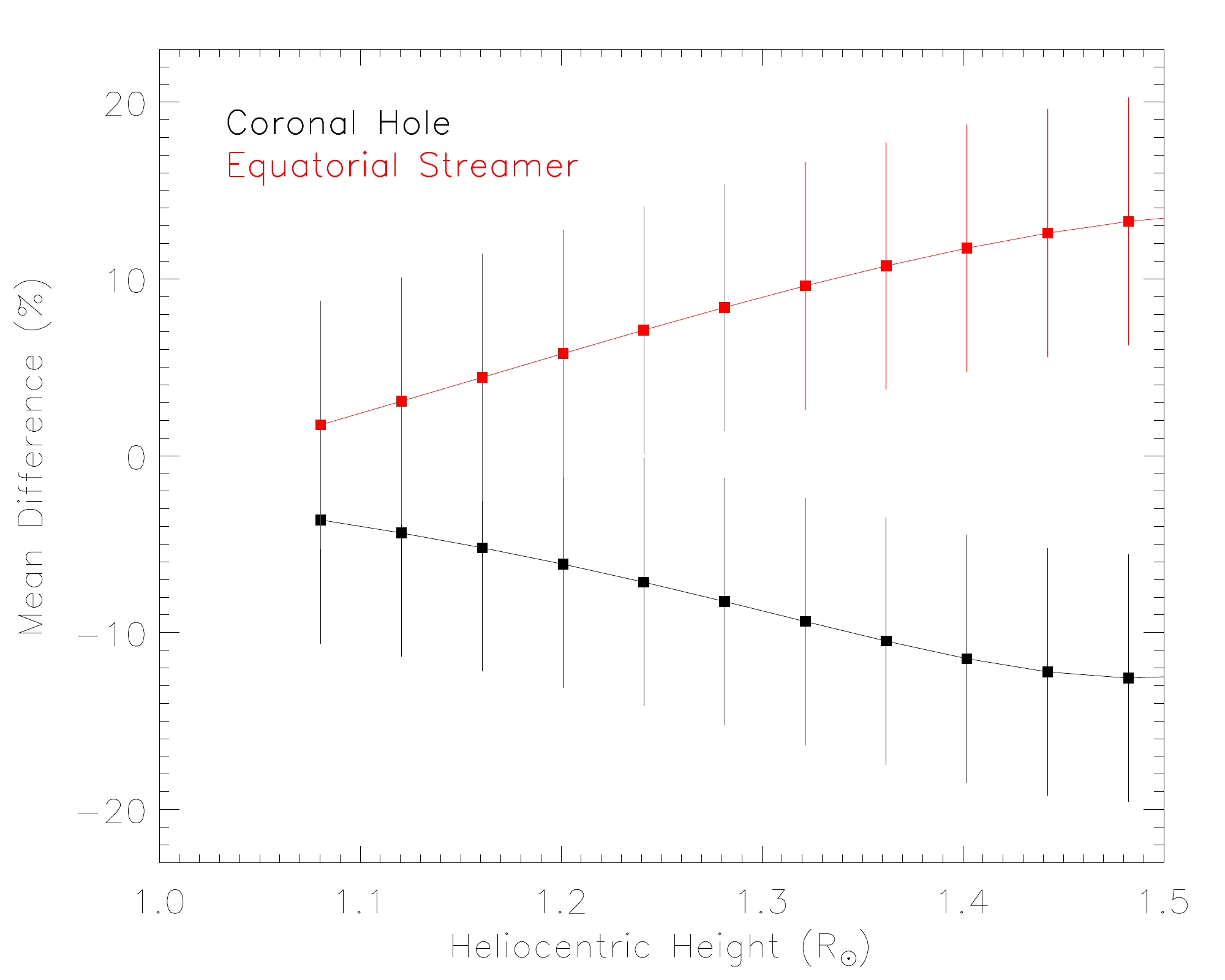}
    \caption{\red Mean difference (expressed as a percentage) between using $pB$ data from Angle Sets A and B for both the coronal hole (black) and equatorial streamer (red) with the error bars representing one standard deviation in both cases}
    \label{fig:ne_difference_comparison}
\end{figure}

Figure \ref{fig:ne_comparison} shows the difference in density between both angle sets A (red) and B (black) as a function of heliocentric height for both the coronal hole (left) and equatorial streamer (right). It is clear to see that the densities found using set A (all polarizer angles) are lower in the coronal hole and higher in the equatorial streamer in comparison to those found using only set B (0\de, 60\de, 120\de). Furthermore, the difference between the two calculated densities clearly becomes greater with increasing heliocentric height, which is shown more clearly in Figure \ref{fig:ne_difference_comparison}. Since the difference increases with increasing heliocentric height, it is unfortunate that the weather conditions on the day of the TSE were sub-optimal because it would be interesting to see how this difference evolves beyond the inner coronal region. However, the existence of this difference implies that changing the number of polarizer angles used does provide a better constraint on the inferred coronal electron densities, particularly for heliocentric heights above $\sim$ 1.5\Rs. For the remainder of the analysis of these results, set A (all polarizer angles) will be used.

\newpage

\subsection{Radial Density Fitting}

\black

Several studies \citep{1977SoPh...55..121S, 1999JGR...104.9801G, 2001ApJ...548.1081H, 2006ApJ...642..523T} state that the radial dependence of the coronal electron density can be expressed in the form of a polynomial:

\begin{equation}
    N_e (r) = \Sigma_i \alpha_i r^{-i}
\end{equation}

where $r$ is given in solar radii and $\alpha$ and $\beta$ are coefficients \red fit to the data. \black Since the analysis of the data obtained in this study \red extended from below $\sim$ 1.1 - 1.5 \black R$_\odot$, three terms is sufficient to provide a good fit to the data, thus, the equation used to fit the data and determine the coefficients was:

\begin{equation}
\label{eq:coefficients}
    N_e(r) = ar^{-b} + cr^{-d} + er^{-f}
\end{equation}

The coefficients for \red both coronal features of interest \black are shown in table \ref{tab:coefficients} and they agree well with previous studies.

\begin{table}[h!]
\caption{Coefficients for equation (\ref{eq:coefficients})}
\label{tab:coefficients}
\begin{tabular}{ccccccc}     
  \hline                   
 & a & b & c & d & e & f \\
  \hline
Coronal Hole & 1.12$\times$10$^6$ & 0.107 & 2.27$\times$10$^8$ & 10.4 & 1.50$\times$10$^8$ & 560 \\
Equatorial Streamer & -2.13$\times$10$^9$ & 6.56 & 1.43$\times$10$^9$ & 6.77 & 1.20$\times$10$^9$ & 6.77 \\
  \hline
\end{tabular}
\end{table}

\section{Conclusion}
\label{sec:conclusion}

The primary aim of this study was to present a \red new design for a lightweight polarization instrument capable of observing using more polarization angles than is typically used, called the Coronal Imaging Polarizer (CIP). The instrument was \black designed and built at Aberystwyth University to observe the polarized brightness of the solar corona during the December 14 2020 TSE for six \red orientation angles of the linear polarizer\black. Due to the design of the instrument, it is very easy to increase or decrease the number of polarization angles depending on the time available during the totality phase of an eclipse. One of the main design elements of CIP was that it would be lightweight in order to be easily transported to an observing site. This element was not only met but was also needed as the team had to relocate to a new observing site on the morning of the eclipse. The new site was a few hours' drive from the original site and with the instrument itself weighing only $\sim$ 1.59 kg, the bulk of the transporting was due to the tripod and tracking mount. Consequently, the entire instrument was easily transported and the simplicity of the design meant that the team could set it up very quickly at the new observing site. However, the weather conditions at the new site still impacted the data - primarily the strong gusts of wind which can be clearly seen in Figure \ref{fig:star locations and drift}, along with airborne dust and high humidity. The instrument was also designed to capture data fully autonomously and efficiently and, again, this aim was met. 

The raw VL images were successfully corrected using flat-field and dark frame subtraction (see Section \ref{subsec:flat/dark correction}) \red and the \black corrected images were then manually coaligned by tracking the drift of a star in the background of the data throughout the duration of totality (see Section \ref{subsec:alignment}). The images were then rebinned to be 8x8 times smaller in order to improve the signal-to-noise ratio which might not be needed in future eclipses given better viewing conditions. The images of different exposure times were combined by using a simple \red signal-to-noise ratio cut \black to create composite images of the eclipse for each individual angle of polarization (see Section \ref{subsec:combination}). These composite images were then combined to give the $pB$ image using a simple least squares fitting method (see Section \ref{subsec:polarization}) \red and relative radiometric calibration was successfully done by cross-calibration with MLSO's K-COR (see Section \ref{subsec:kcor}). The final $pB$ image was then inverted, assuming a locally spherically symmetric corona, to produce radial density profiles for a polar coronal hole and an equatorial streamer. These densities were then compared with previous studies and were found to be in good agreement (see Section \ref{sec:results}). Changing the number of polarizer angles does have an effect on the inferred coronal electron densities, however, due to the inclement weather conditions at the time of observation, no meaningful comparison of the accuracy of more polarizer angles could be performed. It is hoped that CIP can be sent to observe future TSEs to provide better quality constraints on $pB$ and coronal electron densities and build upon the results of this study, particularly with regard to studying the impact of more polarization angles on the quality of TSE data.

\black

\begin{acknowledgements}
    A special note of appreciation must go to the members of the team who braved the COVID-19 pandemic in order to observe this eclipse, without whom this work would not be possible. We acknowledge the valuable advice of the Solar Wind Sherpa team of collaborators led by Prof Shadia Habbal at the University of Hawaii. We also acknowledge studentship funding from the Coleg Cymraeg Cenedlaethol, STFC grant ST/N002962/1 and STFC studentships ST/T505924/1 and ST/V506527/1 to Aberystwyth University which made this instrument and work possible. Some of the $pB$ data and coronal images used in this work are courtesy of the Mauna Loa Solar Observatory, operated by the High Altitude Observatory, as part of the National Center for Atmospheric Research (NCAR). NCAR is supported by the National Science Foundation. \red This research has made use of the Stellarium planetarium. \black
\end{acknowledgements}

\bibliography{references.bib}
\bibliographystyle{aasjournal}

\end{document}